%% file: fst3.tex
%
%
%
%
%
%
%
%
%
%
\newif\ifshowfigs
\showfigstrue
%
%
\font\Cc=cmcsc10

\def\Thsty{\it}
\def\Lesty{\it}
\def\Cl#1{\centerline{#1}}
\baselineskip=15pt
\hfuzz=2pt
\overfullrule=0pt

%
%
%
%
\def\endpage{\vfill\eject}

\def\someroom{\removelastskip\par\vskip 30 true pt 
plus 10 truept minus 10 truept} 
\def\hroom{\hglue 10pt }
\def\pni{\par\noindent}
\def\sni{\someroom\noindent}
\def\leftit#1{\par\noindent\hangindent=40pt\hangafter=1
           \hbox to 30pt{\hglue10pt#1\hss}\hglue 10pt\ignorespaces}
\def\hypit#1{\par\noindent\hangindent=60pt\hangafter=1
           \hbox to 40pt{\hglue5pt#1\hss}\hglue 20pt\ignorespaces}
\headline={\ifodd\pageno\rightheadline \else\leftheadline\fi}
\def\rightheadline{\hfil}
\def\leftheadline{\hfil}
\footline={\ifodd\pageno\rightfootline \else\leftfootline\fi}
\def\rightfootline{\hss\tenrm\folio\hss}
\def\leftfootline{\hss\tenrm\folio\hss}
%
%
\input mnsmac.tex
\edef\bbbQ{\Q}    
\intrmkfalse
\def\ONE{{\tt I}} 
\def\TWO{{\tt II}} 

\def\p{\ve{p}}
\def\q{\ve{q}}
\def\Q{\ve{Q}}
\def\r{\ve{r}}

\def\={=}
\def\half{\sfrac{1}{2}}
\def\set#1#2{\{ #1 : #2\}}

\def\ulj{\underline{j}}

\def\AOne#1{{\sl (H1)$_{#1}$}} 
\def\ATwo#1{{\sl (H2)$_{#1}$}} 
\def\AThr{{\sl (H3)}} 
\def\AFou{{\sl (H4)}} 
\def\AFoup{{\sl (H4')}} 
\def\AFiv{{\sl (H5)}} 
\def\SYmm{{\sl (Sy)}} 
\def\ssni{\par\vskip 15pt \noindent}
\def\half{{\sfrac{1}{2}}}

\def\FOUR{{\tt IV}}

\def\hxp{{h}} 
\def\hxpz{{h_\zer}} 
\def\WL{wicked ladder}
\def\OL{overlapping}
\def\NOL{nonoverlapping}
%
%
\def\socalled{(\TWO.3.4)}
\def\shellj{(\TWO.2.49)}
\def\bestvol{\TWO.1.1}
\def\voilaP{(\TWO.3.5)}
\def\Yuljpi{(\TWO.3.14)}
\def\eazea{(\TWO.3.15)}

\def\goodWL{\TWO.4.8}
\def\bubvol{\TWO.4.7}

\def\zweireg{\TWO.1.2}
\def\YUOm{(\TWO.3.13)}
\def\ImpPow{\ONE.2.35}
\def\Xuljbulnudef{\TWO.3.81}
\def\dreipro{\TWO.3.5}
\def\diffi{(\TWO.3.123)}
%
%
{\nopagenumbers
\rightline{May 1997}
\vglue 2 true cm

\Cl{\bfe Regularity of Interacting Nonspherical Fermi Surfaces: The Exact
Self--Energy}
\vglue 2 true cm
\Cl{\Cc Joel Feldman$^{a,}$\footnote{$^1$}{\rm feldman@math.ubc.ca,
                                        http://www.math.ubc.ca/$\sim$feldman/}, 
Manfred Salmhofer$^{b,}$\footnote{$^2$}{\rm manfred@math.ethz.ch,
                           http://www.math.ethz.ch/$\sim$manfred/manfred.html},
and Eugene Trubowitz$^{b,}$\footnote{$^3$}{\rm trub@math.ethz.ch}}
\vglue 1 true cm
\Cl{$^a$\sl Mathematics Department, The University of British Columbia,
Vancouver, Canada V6T 1Z2}
\Cl{$^b$\sl Mathematik, ETH Z\" urich, CH--8092 Z\" urich, Switzerland}
\vglue 1 true cm
\vglue 1 true cm
\Cl{\bf Abstract}

{\noindent
Regularity of the deformation of the Fermi surface under short-range
interactions is established to all orders in perturbation theory.
The proofs are based on a new classification of all graphs that 
are not doubly overlapping. They turn out to be generalized 
RPA graphs. This provides a simple extension to all orders of the 
regularity theorem of the Fermi surface movement proven in \quref{FST2}. 
Models in which $S$ is not symmetric under the reflection 
$\p \to -\p$ are included.}
\endpage}
%
%
\chap{Introduction}
\pni
This paper is a continuation of \quref{FST1,FST2}, which 
we refer to as \ONE\  and \TWO\ in the following. 
It completes the proof of the regularity of the counterterm function
$K$ that describes the deformation of the Fermi surface under
the interaction. In \TWO, we proved regularity of the 
random--phase--aproximation (RPA) 
self--energy by a detailed, and delicate, analysis of singularities
arising from tangential intersections of the Fermi surface with
its translates. The reason we treated the RPA separately
is not that the RPA is popular or time--honoured,
but that, as we prove in this paper, it
emerges as the leading contribution in a natural way: 
the least regular contributions
to the self--energy are from generalized RPA graphs. 
The contributions from all other
graphs have a higher degree of differentiability. We define
in Section 3.6 what we mean by generalized RPA graphs, but let
us say right away that these graphs occur as natural generalizations
of RPA graphs when the renormalization group scale flow is 
studied. Traditional 
RPA graphs have interaction vertices at scale zero only,
generalized RPA graphs have interaction vertices that are
effective vertex functions of the theory, and lines that 
carry scales. 

We give a very brief outline of our model, and state our hypotheses
and the two main theorems in this Introduction. 
We refer to \ONE\ and \TWO\ for the detailed motivation. 

Our model is a nonrelativistic fermion quantum field theory,
defined by a band structure $e: \cB \to \R, \p\mapsto e(\p)$ 
and an interaction
$\hat v: \R \times \cB \to \C, (p_\zer,\p) \to \hat v(p_\zer,\p)$ 
between the fermions, which has a small coupling constant 
$\la $ in front. 
The set $\cB \subset \R^d$ is a bounded region of momentum space
(for example, the first Brillouin zone) and imposes an 
ultraviolet cutoff on $\p$. 
The function $e(\p)$ is the
energy of a fermion with momentum $\p$ if there is no interaction
between the fermions ($\la=0$),
measured relative to the chemical potential.
In other words, in the system of independent fermions,
and at zero temperature, all states with $e(\p) <
0$ are filled, and all states with $e(\p) >0$ are empty.
The covariance $1/(ip_\zer - e(\p))$, related to the heat kernel
of the free Hamiltonian, 
determines the propagation of free fermions in the system. 
The feature that makes the physical behaviour of the system interesting
and the mathematical treatment difficult is the singularity of
this covariance on the Fermi surface $S=\{\p: e(\p)=0\}$.
The problem addressed in this work is the regularity of the 
movement of the singular set $S$ as the interaction is turned
on.

To state our hypotheses, we need the following standard norms. 
Let $\abs{\cdot}_k$ be the norm 
$$
\abs{f}_k = \sup\limits_{p \in \R\times \cB} \; \sum_{\abs{\al } \leq k}
\abs{D^\al f(p)}
\eqn $$
where  $\al =(\al_\zer, \ldots , \al_d) \in \Z^{d+1}$,
$\al_i \geq 0$ for all $i$, 
is a multiindex, 
$\abs{\al} = \sum\limits_{i=0}^{d} \al_i$, and 
$\del^\al = \left({\del \over \del p_\zer}\right)^{\al_\zer} \ldots 
\left({\del \over \del p_d}\right)^{\al_d}$. 
Let $0 < \hxp \le 1$.
We denote the space of $C^k$ functions on a set $\Om$
whose $k^{\rm th}$ derivative is $\hxp$--H\"older continuous 
by $C^{k,\hxp}(\Om)$, and use the norm
$$
\abs{f}_{k,\hxp} = \sup\limits_{p \in \Om} \; \sum_{\abs{\al } \leq k}
\abs{D^\al f(p)}
+\max\limits_{\al: |\al| =k}
\sup\limits_{x,y\in \Om\atop x\ne y} \frac{\abs{D^\al f(x)-D^\al f(y)}}{\abs{x-y}^\hxp} 
.\eqn $$
For $h=0$, we define $C^{k,0}(\Om)=C^k(\Om)$.

We use the following assumptions on $e$, $\hat v$, and $S$
(not all assumptions are needed in all parts of the proof; 
the details are stated in the Lemmas and Theorems).
For some $\hxp \ge 0$,
\ssni
\hypit{\AOne{k,\hxp}} {\Lesty $\hat v \in C^{k,\hxp}(\R \times \cB,\C)$ 
with all derivatives of 
order at most $k$ uniformly bounded on $\R\times\cB$, and
$\hat v$ satisfies
$\hat v(-p_\zer , \p) = \overline{\hat v(p_\zer,\p)}$.
There is a bounded real--valued function $\tilde v \in C^{k,\hxp}(\cB,\R)$
such that
$\lim\limits_{p_\zer \to \infty} \hat v(p_\zer,\p) = 
\tilde v(\p)$. The convergence is at rate $\abs{p_\zer}^{-\al}$
uniformly in $\p$ for some $\al >0$.
}
\ssni
\hypit{\ATwo{k,\hxp}} {\Lesty$e \in C^{k,\hxp} (\cB,\R)$, and 
$\nabla e(\p)\ne 0$ for all $\p\in S$. }
\ssni
\hypit{\AThr}  {\Lesty The curvature of $S$ is strictly positive everywhere. }
\ssni
\hypit{\AFiv} {\Lesty The Fermi surface $S$ is such that 
$\set{u\p+v\q}{\p,\q\in S, u,v=\pm 1}\subset \openkrnl{\cF}$,
where $\cF$ is the fundamental domain of the action
of the group $\Ga$ of the Fourier space lattice.}
\sni
The meaning of these assumptions is discussed in detail in \ONE\
and \TWO.
We state here only their consequences as regards the constants 
that will appear in the statements of our theorems (for details, see
Lemma \ONE.2.1 and Sections \TWO.2.1 and \TWO.2.2). 
\AOne{} is trivially satisfied if $\hat v$ is a real-valued 
$C^{2,\hxp}$--function that is independent of $p_\zer$,
in other words, if the interaction is instantaneous and 
decays fast enough.
The assumptions \ATwo{2,\hxp} and \AThr\ on $e$ and $S$ imply 
that there is a constant
$r_\zer >0$ such that in the neighbourhood of $S$ given by the
condition $U_{r_\zer} = \{ \p\in \cB: \abs{e(\p)} < r_\zer\}$, 
the following statements hold. 

\leftit{(1)} There is $g_\zer >0$ such that
for all $\p\in U_{r_\zer}$, $|\nabla e (\p)| > g_\zer$. 

\leftit{(2)} 
There is a $C^\infty$ vector field $u$ transversal to $S$, i.e.\
satisfying $|u(\p)| \le 1$ and such that for all $\p \in U_{r_\zer}$,
$|u(\p) \cdot \nabla e(\p)| \ge u_\zer =\sfrac{g_\zer}{2}$,
and there is a $C^{2,\hxp}$--diffeomorphism $\ph: ((-r_\zer,r_\zer)
\times S^{d-1} \to U_{r_\zer} \subset \cB$ such that 
$e(\ph(\rh,\th))=\rh$. Furthermore, when $d=2$, 
$\del_\th\ph(0,\th)$ is of constant
(nonzero) length. Throughout, we denote the function $\ph$  by $\p$, 
i.e., we write $\p(\rh,\th)=\ph(\rh,\th)$. $\rh$ and $\th$ are
our standard radial and tangential coordinates around $S$.

\leftit{(3)}  
There is a constant $w_\zer>0$, related to the
minimal curvature $\ka_\zer$ in \AThr, such that, e.g. in $d=2$,
for all $\p \in U_{r_\zer}$, $(\del_\th \p, e''(\p) \del_\th\p)
\ge w_\zer$ (here $e''$ is the matrix of second derivatives of
$e$).

For the full regularity proofs, we need another hypothesis.
We call a band structure $e$ {\it symmetric} if
\ssni
\hypit{\SYmm} {\Lesty For all $\p \in \cB$, $e(-\ve{p} ) = e(\ve{p} )$.}
\ssni
and {\it asymmetric} otherwise. 
We require that either $e$ is symmetric, or, if $e$ is asymmetric, 
we assume \AFou\ and \AFoup\ (stated in Chapter 2 of \TWO).
We do not repeat these assumptions in all detail here, 
since their main consequences were already derived in \TWO.
They concern the curvature of
$S$ at a point $\p \in S$ and at its antipode $\ve{a}(\p) \in S$
(defined by the requirement that the unit normal $\ve{n}$ 
at $\ve{a}(\p)$ should be the opposite to that at $\p$,
$\ve{n} (\ve{a}(\p))=-\ve{n}(\p)$). 
\AFou\ requires that these two curvature must not differ by too
much (if \SYmm\ holds, they are equal), where `too much' means
a fixed number. \AFoup\ requires, on the other hand, that for
asymmetric $e$, the curvature at a point and its antipode
have to differ at all but a finite number of points. These geometrical
conditions on the Fermi surface lead to rather delicate bounds
which imply, for the models with an asymmetric $e$,
the regularity of the RPA self--energy and counterterm,
as well as the suppression of the Cooper instability
(see \TWO, Appendix C and Theorem \TWO.4.9). As discussed in
\TWO. the regularity proof is easier for symmetric $e$, partly
because the antipode function is, by definition, in general 
only $C^1$ if $e$ is $C^2$, and partly because certain singular
points arise only in the asymmetric case (see Lemma \TWO.4.6).

The connected, amputated Green functions of the interacting system
are generated by the effective action. Because of the singularity,
we regularize the propagator by a cutoff $\veps=M^I$ (here $M>1$
and $I\in \Z, I<0$ (this will be recalled in detail in Chapter
3) and study the limit $\veps \to 0$. 
At positive $\veps$, the effective action 
$$
e^{\cG(\ch,\chq)} = \int d\mu_{C(\veps)}(\ps,\psq) e^{\la V^{(0)} (\ps+\ch,\psq+\chq)}
,\eqn $$
with $C(\veps)$ the regularized covariance and $V^{(0)}$ the
initial interaction given by $\hat v$ (see \ONE.1.23),
is well-defined.
For $\veps >0$, an expansion of the exponential in $\la$, the
polymer expansion for the logarithm, and a determinant bound,
imply that $\cG$ is analytic in $\la $ in a disk whose radius
is $\veps$--dependent. As $\veps \to 0$, the radius of convergence
shrinks to zero, and even the coefficients of $\la^r$ diverge
for all $r \ge 3$ (`infrared divergences'). 

In \ONE,\TWO, this paper, and a forthcoming paper (\FOUR), 
we prove the existence of
the limit $\veps \to 0$ of the effective action in any finite
order in perturbation theory in the coupling $\la$. 
As explained in \ONE\ and \TWO, 
this is a rather nontrivial problem even at the perturbative
level, because in the absence of spherical symmetry
(which is always broken by the crystal lattice) one has to do
renormalization with momentum--dependent counterterms.
Essentially, one has to prove regularity of the Fermi 
surface deformation under the interaction. It is
the purpose of the present series of papers to show that this
is indeed possible. The counterterm function $K$ depends on 
$e$, $\la V$, and $\p$, $K=K(e,\la V,\p)$. When the counterterm
$(\bar\ps, K \ps)$ is added to the action, the Fermi surface stays
fixed, independent of $\la$, 
because the counterterm removes those parts of the self-energy
that lead to the movement and deformation of the Fermi surface.
Consequently, as shown in \ONE, the infrared divergences of the
unrenormalized expansion disappear. 
In reality, however, the counterterm is not there at the beginning,
and the Fermi surface responds to the interaction by a deformation.
Merely adding a counterterm changes the model.
It is explained in detail in \ONE\ and \TWO\ that to renormalize
without changing the model, one has to solve the equation
$$
E=e+K(e,\la V)
\EQN\Wltfrml $$
in a suitable space $\cE$ of functions from $\cB$ to $\R$. 
The results of \TWO\ and the present paper imply, very roughly speaking,
that $\cE$ can be chosen as  a ball of fixed radius in 
$(C^2(\cB,\R),\abs{\cdot}_\two)$ around a given $E$ which gives
rise to a $C^2$ Fermi surface with strictly positive curvature.
More precisely, we have the following theorem about the formal
power series for $K$,
$$
K(e,\la V,\p) = \sli_{r=1}^\infty \la^r\; K_r(e,V,\p)
,\eqn $$
and, for $d\ge 3$, the self energy
$$
\Si (p) = \sli_{r=1}^\infty \la^r\; \Si_r (p)
\eqn $$

\The{\Kalloreg}{\Thsty 
\leftit{$(i)$} Let $d=2$. For
all $\hxp \in [0,\half)$: if \AOne{2,\hxp}, \ATwo{2,\hxp},
\AThr, \SYmm, and \AFiv\ hold, then there is a constant
$\cC_\one$ such that for all $r \ge 1$, $K_r \in C^{2,\hxp}(\cB,\R)$,
and
$$
\abs{K_r}_{2,\hxp} \le {\cC_\one}^r \; r!
\eqn $$
The constant $\cC_\one$ depends only on $\hxp$,
$\abs{e}_{\two,\hxp}$, $\abs{\hat v}_{\two,\hxp}$, $g_\zer$, $r_\zer$, and 
$w_\zer$. $\hxp=0$ is allowed. 

\leftit{$(ii)$} Let $d=2$. For
all $\hxp \in (0,\sfrac{1}{3})$: if \AOne{2,\hxp}, \ATwo{2,\hxp},
\AThr, \AFou, \AFoup, and \AFiv\ hold, then there is a constant
$\cC_\two$ such that for all $r \ge 1$, $K_r \in C^{2,\hxp}(\cB,\R)$,
and
$$
\abs{K_r}_{2,\hxp} \le {\cC_\two}^r \; r!
\eqn $$
The constant $\cC_\two$ depends only on $\hxp$,  $\abs{e}_{\two,\hxp}$, 
$\abs{\hat v}_{\two,\hxp}$, $g_\zer$, $r_\zer$, $K_a$
and $w_\zer$. $\hxp=0$ is not allowed. ($K_a$ is defined in \AFoup).

\leftit{$(ii)$} Let $d\ge 3$. There is $\hxpz>0$ such that for
all $\hxp \in [0,\hxpz]$: if \AOne{2,\hxp}, \ATwo{2,\hxp},
\AThr, and \AFou\ hold, then there is a constant
$\cC_\thr$ such that for all $r \ge 1$, $K_r \in C^{2,\hxp}(\cB,\R)$,
and
$$
\abs{K_r}_{2,\hxp} \le {\cC_\thr}^r \; r!
\eqn $$
The constant $\cC_\thr$ depends only on $d$, $\hxp$, $\abs{e}_{\two,\hxp}$, 
$\abs{\hat v}_{\two,\hxp}$, $g_\zer$, $r_\zer$,
and $w_\zer$. $\hxp=0$ is allowed. Moreover, there is a constant
$\cC_4$ such that for all $r \ge 1$, $\Si_r \in C^{2,\hxp}(\cB,\R)$,
and
$$
\abs{\Si_r}_{2,\hxp} \le {\cC_4}^r \; r!
.\eqn $$

}

\sni
For an iteration of the map to get a solution,
as done in \FOUR, we need the constants to be uniform on the
set $\cE$ where we want to do the iteration. 
The quantities $g_\zer$, $r_\zer$ and $w_\zer$ are
all uniform even for $e \in C^{2,0}$, i.e., with $\hxp=0$. 
The set $\cE$ will be defined, and the iteration will be done, 
in \FOUR.

To prove Theorem \Kalloreg, we use the Feynman graph representation,
which is a rewriting of every $K_r$ (and every $\Si_r$) as 
a sum over values of graphs with $r$ vertices. 
A restriction of Theorem \Kalloreg\ to the contributions from
RPA graphs was proven in \TWO. Here we extend it to the 
full $K$. 
We also prove the following statement about the self--energy,
announced in \TWO, which is much simpler since it 
requires only an analysis of simply overlapping graphs.
It holds without the filling restriction \AFiv, and, for 
asymmetric $e$, without the condition \AFoup.

\The{\Salloreg}{\Thsty Let $d\ge 2$ and $0 \le \ga < 1$.
If \AOne{1,\ga}, \ATwo{2,0}, \AThr, and \AFou\ hold, then
$\Si_r \in C^{1,\ga} (\R\times \cB,\C)$ for all $r \ge 1$.
In particular, if \AOne{2,0}, 
\ATwo{2,0}, \AThr, and \AFou\ hold, then
$\Si_r \in C^{1,\ga} (\R\times \cB,\C)$
for all $\ga < 1$ and all $r\ge 1$. }

\sni
As noted in \TWO, the calculations in \quref{F} indicate that
in two dimensions,
$\Si_2$ is not twice differentiable because of a logarithmic
singularity in the second derivative, even for an $e\in C^\infty$ 
and $\hat v=1$. As discussed in detail
in Section 1.3 of \TWO, this is the main reason why we take
the function $K$, and not $\Si$ itself, to do renormalization.
Note also that while $\Si_r \in C^{1,\ga}$ requires only 
$\hat v \in C^{1,\ga}$, we have to require $e$ to be at least 
$C^{2,0}$ because our proof of the volume bound 
Theorem \TWO.1.1, on which Theorem \Salloreg\ is based,
requires $e \in C^2$.

As stated in Theorem \Kalloreg, in three or more dimensions, 
not only $K$, but also $\Si $ is a $C^2$ function of its arguments.
Although a superficial analysis would suggest that the power counting
behaviour is essentially independent of dimension because
the codimension of the Fermi surface $S$ is one in any dimension
$d$, the truth is that there is a nontrivial dimension-dependence
even in perturbation theory.  
In our case, the main reason for the better behaviour
in $d \ge 3$ is that the singularities of the Jacobian analyzed
in \TWO\ are point singularities on the strictly convex Fermi
surface, i.e.\ point singularities located on a $(d-1)$--dimensional
submanifold. The codimension of these point singularities on
$S$ is $d-1$ and thus grows with $d$, and, as shown in \TWO, 
Section 3.5, this implies the better regularity properties of $\Si$. 

The analysis done in \TWO\ was rather intricate even for the
very small class of RPA graphs, and a generalization of that
analysis to all graphs contributing to $K$ looks very complicated.
The observation which allows us to avoid having to extend the
analysis of \TWO\ to all graphs is the following.
Each volume gain from overlapping loops increases the degree
of smoothness by almost one derivative. 
If there were no volume gains,
graphs contributing to the proper selfenergy would be $C^{0,\ga}$
for every $\ga < 1$. 
Thus the value of any graph that provides two volume gains is
$C^{2,\ga}$, hence almost three times differentiable.
We call such graphs doubly overlapping.
The heart of this paper is the classification of all graphs that
are not doubly overlapping. They turn out to be the generalized
RPA graphs mentioned above. The analysis of \TWO\ applies to
these graphs, so a combination of the results of \TWO\ with the
bounds for values of doubly overlapping graphs (along the lines
of the tree decomposition done in \ONE) implies Theorem \Kalloreg.

It is a natural question whether the statement that $e$ is $C^2$
implies that $K$ is $C^2$ generalizes to a statement $e\in C^k$
implies $K \in C^k$, at least if $e$ is symmetric. In \TWO, we
proved this for the class of RPA graphs. As mentioned, an extension
of this proof looks rather difficult, even when combined with
further volume techniques. 

We end with an overview of the rest of the paper. Chapter 2 
contains the main new idea of the proofs given here. It is
graph theoretical and can therefore be understood without any
knowledge of the scale decomposition. All nontrivial graph theoretical notions
required are defined there, so this part is self--contained.
In Chapter 3 we prove Theorems \Kalloreg\ and \Salloreg.
We outline the strategy at the beginning of Chapter 3.
Although the idea of the proof will be clear from the developments
in Chapter 2, familiarity with the results of \ONE\ and \TWO\ is
required for an understanding of the details. It is a good idea
to keep a copy of \ONE\ and \TWO\ in reach when reading Chapter
3 because we shall refer frequently to these
papers, and use the notation and results thereof.

\chap{Classification of Skeleton Graphs}
\noindent
We denote the vertex set of a graph $G$ by $V(G)$ and its 
set of internal lines by $L(G)$. 
A line is internal (resp.\ external) if both (resp.\
only one) of its ends are hooked to vertices (resp.\
is hooked to a vertex).
If a vertex $v \in V(G)$
has incidence number $n$, we call it an $n$-legged vertex. 
A vertex is called external if it is hooked to an external
leg of the graph. 
If all vertices of $G$ have even 
incidence number, then $G$ has an even number of external legs. 
We assume throughout that all graphs have vertices with even 
incidence number. This is the case for the effective vertices
of our model. Thus all graphs (and subgraphs) appearing in our
analysis have an even number of external legs.

We define a skeleton graph to be a connected graph with at least two vertices,
that has no two--legged vertices and no proper two--legged subgraphs. 
If $G$ is a two--legged skeleton graph, $G$ is one-particle irreducible
(1PI). 
That is, $G$ cannot be disconnected by cutting one internal line.
In fact, we show in Section 2.3 the stronger statement
that, if all vertices of the two--legged graph $G$ have 
even incidence number then $G$ is a skeleton graph
if and only if $G$ is two--particle--irreducible. 

\sect{Doubly overlapping graphs}
\noindent
We briefly recall the notion of overlapping loops defined in \ONE.
Let $T$ be a spanning tree for $G$, i.e.\ a subgraph of $G$ that
is a connected tree and contains all vertices of $G$, 
and $\ell \not\in L(T)$ a line
of $G$. 
We associate to $\ell$ a loop in $G$ as follows. 
Denote the vertices at the ends of $\ell $ by $v$ and $w$.
If $v=w$, the loop generated by $\ell $ contains only the line $\ell $. 
If $v\ne w$, there is a unique path $P_\ell$ in $T$ from $v$ to $w$. 
The loop generated by $\ell $ (and $T$)
 contains $\ell$ and all lines of $P_\ell$.
A graph is \OL\ if for some choice $T^*$ of the spanning tree there are 
lines $\th\in L(T^*)$ and $\ell_\one \ne \ell_\two \in L(G)\setminus L(T^*)$
such that the loops generated by $\ell_i$ both contain $\th$
(this statement is equivalent to the definition given in \ONE, 
see Remark \ONE.2.18 $(iv)$). It was shown in Lemma \ONE.2.34, 
that if $G$ is \OL, then not only $T^*$, but every spanning tree 
of $G$ has this property. It is straightforward to verify that the `sunset' 
graph shown in Figure \nextfig{\dolone} is 
\OL\ according to the above criterion. 
More generally,
any graph that contains a sunset subgraph is \OL. 
We showed in Section 2.4 of \ONE\ that the only \NOL\ graphs
with two external legs are generalized Hartree-Fock graphs
and that the only \NOL\ graphs with four external legs are 
the dressed bubble chains which contribute to a generalized RPA
resummation. 

\herefig{fst3fig00}

\Def{\DOLDef}
We say that a spanning tree $T^*$ gives rise to two separate overlaps
if there are lines $\th\ne \ze \in T^*$ and 
$\ell_\one,\ell_\two,k_\one,k_\two\in L(G)\setminus L(T^*)$, all distinct, 
such that the loops generated by $\ell_\one $ and $\ell_\two$ both 
contain $\th$, and the loops generated by $k_\one$ and $k_\two $ both 
contain $\ze $, but not $\th$. 
We call a graph doubly overlapping (DOL) if it has a spanning tree $T^*$ 
that gives rise to two separate overlaps.

\ssni
Since by this definition, a DOL graph has to have at least four loops,
the sunset graph of Figure \dolone\ is not DOL. Moreover, the
graph has to have at least three vertices, because otherwise
there can be at most one line in the spanning tree.
By the usual counting arguments, a connected graph with 
$E=2$ external legs and only four-legged vertices has 
$|V(G)|$ loops. Thus any such graph with at least four vertices
has at least four loops. For any such graph to be DOL, 
it must have at least four vertices. Replacing a four-legged
vertex by one with more than four legs increases the number of
loops. There exists a two-legged graph with three vertices,
one of which is six-legged, that is DOL (see Figure 3).
 
The significance of the notion of overlapping graphs for our
analysis is that the value of any \OL\ graph, 
all of whose lines are restricted to carrying momenta $p$
obeying $|e(\p)| \le \veps$, contains a subintegral 
bounded by 
$$
\cW(\veps ) = \sup\limits_{\q \in \cB}
\max\limits_{v_i \in \{ \pm 1\}} 
\int\limits_{S^{d-1}\times S^{d-1}} d\th_\one \, d\th_\two \,
\True{\abs{e(v_\one \p(0,\th_\one) + v_\two \p (0,\th_\two ) + \q )} \leq \veps}
.\EQN\cWdef $$
Here $\p(\rh,\th)$ denotes a parametrization of a neighbourhood of the Fermi 
surface $S$ with $\rh=0$ corresponding to the Fermi surface itself
(see Section 2.2 of \TWO). By Theorem \bestvol, there is a constant
$Q_V$, depending only on $\abs{e}_{2,0}$, $r_\zer$, $g_\zer$,
and $w_\zer$, such that
$$
\cW(\veps) \le Q_V \veps \abs{\log \veps}
.\eqn $$
This {\it volume improvement bound} leads to an improvement in power counting,
as proven for all \OL\ graphs in Lemma \ONE.2.35
(and as discussed once more in detail in second order in \TWO),
which implies differentability of the self-energy
and the counterterm function. 

We show below that the spatial momentum integrals
for any DOL graph with lines of energy scale $\veps $ contain an improvement 
factor $\cW(\veps)^2$, where $\cW$ is the volume improvement function
defined in \queq{\cWdef}, and that this implies that the value
of all DOL graphs is $C^2$. 

\Rem{\quoDOL} If $G$ has a connected subgraph $H$ such that $H$ and 
$G/H$ are \OL, then $G$ is DOL. 

\Proof Recall that $G/H$ is the quotient graph of $G$ obtained by 
replacing $H$ by a vertex, so that in particular $L(G/H)=L(G)\setminus L(H)$. 
We shall show that under the given hypotheses, any choice of the 
spanning tree for $G$ that remains a tree in $G/H$ 
gives rise to two separate overlaps. Let $T$ be such a spanning tree for $G$. 
Then its restriction to $H$, $T_\one$, is a spanning tree for $H$, 
and its quotient $T_\two$ is a spanning tree for $G/H$. 
$T_\one$ only consists of lines internal to $H$. 
$H$ is \OL, so there are two lines $k_\one\ne k_\two
\in L(H)\setminus L(T_\one)$ such that the loops generated 
by $k_\one$ and $k_\two$
contain a common line $\ze \in L(T_\one)$. Both loops never get out of $H$, so 
they contain no line of $L(G)\setminus L(H) = L(G/H)$, 
and therefore no line of $T_\two$.  
Since $G/H$ is \OL, there are two lines $\ell_\one \ne\ell_\two$ in 
$L(G/H)\setminus L(T_\two)$ whose loops contain a line $\th\in L(T_\two)$.
We now turn to the situation in $G$ itself. The loops generated by 
$\ell_\one, \ell_\two$ in $G$ still contain $\th$. If these loops
contain lines of the subgraph $H$, they may also contain the line $\ze$. 
But the loops generated by $k_\one$ and $k_\two$ remain unchanged, 
so they contain $\ze$, but not $\th$. Thus $G$ is DOL.
\endproof

\Rem{\tjaja} Not every spanning tree for a DOL graph $G$ gives rise 
to two separate overlaps. An example is given in Figure \nextfig{\figutree}.
The heavy lines are those in the spanning tree.

\herefig{fst3fig01}

\ssni
If $G$ is two-legged, 1PI, and has two external vertices,
then $G$ is overlapping by Lemma \ONE.2.22. 
There are paths of minimal length between these two vertices.
Let $\th$ be one of them, identify the map $\th$ with the 
subtree of $G$ that it defines, let $\abs{L(\th )}= t $
and number the vertices of $\th$ in the order of the walk from 
the first to the second external vertex as 
$v_\zer , \ldots , v_t $. The integer $t$ is the minimal number of steps 
required to walk from one to the other external vertex.

\The{\skeledol}{\Lesty All two--legged skeleton graphs with $t\ge 2$ are doubly
overlapping.}
\sni
This theorem is a direct consequence of the following, more detailed, lemma.  

\Lem{\dblegain} { \Lesty 
Let $G$ be a two--legged skeleton graph and $t \geq 2 $. Then $G$ is DOL.
More precisely, for all $r,s$ with $0 \leq r < s \leq t-1$,  
$G$ takes the form shown in Figure \nextfig{\figufiv}. 
The subgraphs $G_\one $ and $G_\two $ are connected, and 
each of them connects to $G'$ by at least three lines.
If it is not possible to choose 
$G_\one$, $G_\two$ and $G'$ all connected,
then  \senzafig{\figusix}
$G$ is as indicated in Figure \nextfig{\figusev}, with $C'$ and $C_\one$
connected and $\overline{m}_\one \ge 2$ ($\overline{m}_\one$ is the number 
of lines joining $G_\one$ to $C'$) and $n_\one \ge 3$,
or as in Figure \nextfig{\figueig}, with $C_\one$ connected, and with $\overline{m} \ge 2$ 
and $\overline{n} \ge 2$. 

}

\Proof For $v \in V(G)$, let $s(v) $ be the minimal number of 
steps required to walk from $v_\zer $ to $v$ over lines of $G$.
Thus $s(v_\zer )=0$, and $s(v_r) = r$ for all $r \in \nat{t}$
by minimality of $\th$
(of course, there may also be $v \in V(G)$ for 
which $s(v)> t$). 
Let $S_k$ be the subgraph of $G$ defined as follows. 
The set of vertices is 
$V(S_k) = \{ v \in V(G): s(v) \leq k \}$.
The lines of $S_k$ are all those 
lines $l \in L(G)$ that join vertices in $S_k$.
Obviously, $S_k$ is a connected subgraph of $G$ 
that contains $v_\zer$. If $k=0$, $S_k$ is the 
vertex $v_\zer $. 
Also, $v_k \in S_k$ and $v_{k+1} \not\in S_k$.
Let $T_m$ be the analogous graph constructed from 
$v_t$, i.e. if $s'(v) $ is the minimal number of steps 
from $v_t$ to $v$, going over lines of $G$,
$V(T_m) = \{ v \in V(G): s'(v) \leq m\}$.
$T_m$ is a connected subgraph of $G$ that contains $v_t$,
and $v_{t-m} \in T_m$, but $v_{t-m-1} \not\in T_m$.

\herefig{fst3fig02}

For $r$ and $s$ as given in the statement of the Lemma, 
let $G_\one =S_r$ and $G_\two = T_{t-s-1}$; both of these
graphs are nonempty for all $t\ge 2$ (if $t=2$, $G_\one=v_\zer$ 
and $G_\two=v_\two$). Moreover, they are disjoint
by minimality of $\th$ and because $r + t-s-1 < t-1$. The vertices
$v_{r+1}$ and $v_s$ are in neither of the two graphs
($ v_{r+1}=v_s$ is possible).
Let $G'$ be the subgraph of $G$ with vertex set
$V(G') = V(G) \setminus (V(G_\one) \cup V(G_\two))$
and with those lines of $G$ that join vertices in $V(G')$. 

The remaining lines $l \in L(G) \setminus 
(L(G') \cup L(G_\one) \cup L(G_\two)$ are external 
lines of the three subgraphs, and they connect $G'$, $G_\one$ and
$G_\two$ to form $G$. However, by minimality
of $\th$, there can be no line that connects a vertex
in $G_\one $ to a vertex in $G_\two$ (if there were 
such a line, it could be used to make a path of length
strictly less than $t$ between $v_\zer $ and $v_t$). 

All vertices of $G$ have an even number of legs, so 
$G_\one $, $G_\two$ and $G'$ must all have an even 
number of legs. Thus, the number of lines $k_\one$ between 
$G_\one $ and $G'$ is odd, and the number of lines $k_\two$ 
from $G'$ to $G_\two$ is odd. 
$G$ is 1PI, so $k_\one \ge 3$ and $k_\two \ge 3$,
hence $G$ is of the form shown in Figure $\figufiv$.

If $G'$ is disconnected, we 
decompose it into its connected components $C_\al$.
Let $C$ be one of these components. If all external lines
of $C$ connect to $G_\one$, an even number of external 
lines of $G_\one $ is bound. We absorb all these components 
in $G_\one$. Similarly, we absorb in $G_\two$ all $C_\al$'s
connected directly only to $G_\two$. 
There is at least one component of $G'$ 
that connects to both $G_\one$ and $G_\two$. If there is only one
such component, then $G$ is as in Figure $\figufiv$. 
Otherwise, $G$ is as shown in Figure 
$\figusix$, where $a \ge 2$, and
all subgraphs $G_\one$, $G_\two$ and $C_\one, \ldots, C_a$ are 
connected. Since all the subgraphs have an even number of legs,
and because $G$ has no two--legged subdiagrams, 
$$
m_b\ge 1, \quad n_b\ge 1, \quad  m_b+n_b \hbox{ is even and }
m_b+n_b \ge 4
\EQN\numbertheory $$
holds for all $b\in \{ 1,\ldots,a\}$. Moreover, 
$$
\sum\limits_{b=1}^a m_b \quad \hbox{and}\quad
\sum\limits_{b=1}^a n_b \quad \hbox{ are odd} 
.\EQN\itsodd $$
Note that the $m_b$ and $n_b$ do not both have to be $\ge 2$; the dots 
in Figure $\figusix$ should be interpreted that way. 

\herefig{fst3fig03}

Assume that there is $k$ such that $m_k=1$ or $n_k=1$. By the symmetry 
of the graph and by renumbering, we may assume that $m_\one=1$.
Then $n_\one\ge 3$ by \queq{\numbertheory}.
We redraw $G$ by collecting $C_\two, \ldots, C_a$ and $G_\two$ in
a subgraph $C'$. $C'$ is connected, and $G$ takes the form shown in 
Figure $\figusev$. $\overline{m}_\one = \sum\limits_{b=1}^a m_b -1$ is even by 
\queq{\itsodd} and nonzero because $G$ is 1PI, hence $\overline{m}_\one \ge 2$. 

If $m_b>1$ and $n_b>1$ for all $b \in \{ 1,\ldots, a\}$, there is a
$k$ such that $m_k$ is odd and hence at least $3$.
 Without loss of generality, we may assume
that $k=1$. By \queq{\numbertheory}, $n_\one$ must also be odd,
hence $n_\one \ge 3$ as well. We collect $C_\two, \ldots, C_a$ into 
a subgraph $D$. $D$ is disconnected if $a>2$, but always nonempty.
This brings $G$ into the form shown in Figure $\figueig$. By construction of 
$G_\one$ and $G_\two$, and because $D$ is nonempty, $\overline{m}>0$ and
$\overline{n}>0$. By \queq{\itsodd}, they must be even, so 
$\overline{m}\ge 2$ and $\overline{n}\ge 2$.

In Figures $\figufiv $, $\figusev $, and $\figueig $, 
the lines drawn fat are those in a possible 
spanning tree, where one can see directly from the definition 
that the graph is DOL (in all cases, one can choose a quotient 
graph $H$ such that Remark \quoDOL\ holds, e.g.\ in Figure $\figusev$, 
$H$ can be taken to consist of $C_\one$ and $C'$ and the lines joining them).
\endproof

\herefig{fst3fig04}

\herefig{fst3fig05}

\def\MS{multiple sunset}
\sect{Graphs with $t=1$}
\noindent
To complete the classification of doubly \OL\ graphs, we turn to the case 
$t=1$. We call the external vertices $v_\zer $ and $v_\one$. 
If $G$ has only these two vertices, then $G$ is the sunset diagram shown in 
Figure $\dolone$, or $G$ is the \MS\ shown in Figure \nextfig{\figunin}. 
Otherwise, let $D$ be the subgraph of $G$
consisting of all those vertices of $G$ other than the external
vertices, and of all the lines between these
vertices. The connected components of $D$ may be joined only to $v_\zer$,
only to $v_\one$, or to $v_\zer $ and $v_\one$. All components that 
connect only to $v_\zer $ are absorbed in a connected subgraph 
$G_\one$ of $G$, and similarly we construct $G_\two$. If there is no 
component of $D$ that connects both to $v_\zer$ and $v_\one$, then 
$G$ is as shown in Figure $\dolone$ or $\figunin$, 
with the disks now representing the subgraphs $G_\one$ and $G_\two$. 
If $D$ is nonempty and $G_\one$ has at least one vertex in addition to
$v_\zer$, then, by the definition of skeleton graphs, 
$G_\one$ is \OL\ and hence is $G$ is DOL. 

\herefig{fst3fig06}

\noindent
Otherwise, $G$ is as in Figure $\figusix$, except that now there are $m\ge 1$ 
additional lines that join $G_\one$ directly to $G_\two$. We first show 
that for $a\ge 2$ components $C_b$ between $G_\one$ and $G_\two$, the
graph is DOL. Since adding lines to a DOL graph keeps it DOL, 
it suffices to consider the `minimal' case $a=2$, $m=1$, and $m_\one+n_\one=
m_\two + n_\two =4$. Up to exchanges of $G_\one$ and $G_\two$,
or of $C_\one $ and $C_\two$, there are only the four cases shown in 
Figure \nextfig{\figuten}. They are all DOL by direct inspection.

\herefig{fst3fig07} 

If $a=1$, we consider $m\ge 2$ first. Again, we may reduce to the minimal
possible number of lines, $m_\one+n_\one=4$, $m=2$ (reduced from general
$m$ even), or $m=3$ (reduced from general $m$ odd). These graphs are drawn in Figure 
\nextfig{\figuele}; they are
DOL.

\herefig{fst3fig08}

There remains the case $m=1$. If $m_\one+n_\one\ge 6$,  we may reduce to 
$m_\one+n_\one=6$. Then $G$ looks as in 
Figure \nextfig{\figutwe} $(a)$, so it is DOL. 
If $m_\one+n_\one = 4$, $G$ is as in Figure $\figutwe$ $(b)$.
\herefig{fst3fig09}
If any of the subgraphs drawn as shaded disks is \OL, $G$ is DOL. 
If the subgraphs are \NOL, they are dressed bubble chains (see 
Definition 2.24 and Figure $\figunin$ in \ONE) by Lemma 2.26 of \ONE. Since 
$G$ has $t=1$, $G$ must then look as in Figure \nextfig{\figuthi},
with the disks representing connected subgraphs,
\senzafig{\wicked}
or as in Figure $\wicked$, with the disks representing either
four-legged vertices or 
$2m$--legged vertices with $m-2$ self--contractions.

\herefig{fst3fig10}

\herefig{fst3fig11}
The graph in Figure $\figuthi$ is DOL, 
but the one in Figure $\wicked$, the \WL, is not. 
In summary, we have proven the following theorem.

\The{\allnonDOL}{\Lesty Let $G$ be a non--DOL two--legged skeleton graph. 
Then $G$ is a sunset, or a \MS, or a \WL, with the vertices possibly being 
$2m$--legged vertices with $m-2$ self--contractions.}

\sect{Two--particle irreducible graphs}
\noindent
A connected graph is called two--particle irreducible (2PI) if
cutting any two particle lines does not disconnect the graph
and two--particle reducible (2PR) otherwise.
In this section we give a characterization of 2PI graphs with
two and four external legs. The statement that $G$ has a two--legged
subgraph $H$ includes the requirement that 
$H$ is a proper subgraph of $G$.

\Pro{\twoPI}{\Thsty Let $G$ be a 1PI graph with vertices that
all have even incidence number. 
\leftit{$(i)$} Let $G$ be two--legged. Then 
$$
G \hbox{ is 2PR } \Longleftrightarrow G \hbox{ has a two--legged
subgraph}
\eqn $$
In other words, all the two--legged 2PI graphs are skeleton graphs,
and vice versa. 
\leftit{$(ii)$} Let $G$ be four--legged. Then 
$$\eqalign{
G \hbox{ is 2PR } \Longleftrightarrow \quad & G \hbox{ has a two--legged
subgraph or} \cr
& G \hbox{ is as shown in Figure \nextfig{\doltwoPI}}, \cr
& \hbox{where $C_\one$ and $C_\two$ are 1PI}
\cr}\eqn $$

}
\herefig{fst3fig12}
\Proof For a graph $G$ and  line $l$ of $G$, $G-l$ denotes the
graph where $l$ is cut. $E(G)$ denotes the number of external
legs of the graph $G$.

It is obvious that if a graph $G$ contains
a two--legged subgraph $H$, $G$ is 2PR since cutting the lines that connect
$H$ to the other vertices of $G$ makes the remaining graph
disconnected. It is also obvious that the graph shown in 
Figure $\doltwoPI$ is 2PR. So $\Leftarrow$ holds, and it remains
to prove $\Rightarrow$. 

Let $G$ be 2PR. 
Let $l_\one$ and $l_\two$ be any two lines such that $G-l_\one-l_\two$
is disconnected. Since $G$ is 1PI, $G-l_\one$ and $G-l_\two$ are
both connected, so $G-l_\one-l_\two$ can fall into at most two
connected components, $C_\one$ and $C_\two$, which
are joined by $l_\one$ and $l_\two$ only,
since otherwise $G-l_\one-l_\two$ would be connected. 
Cutting a line gives two external legs, so 
$$
E(C_\one)+E(C_\two) = 4 + E(G)
\eqn $$
must hold. 

Let $E(G)$ =2.  
There are two cases: $(a)$ If $C_\one$ and $C_\two$ 
both contain an external leg of $G$, $G$ is as shown in Figure \nextfig{\doltwoL}
$(a)$. $(b)$ If only one of them connects to an external leg, $G$ is
as shown in Figure $\doltwoL$ $(b)$, and thus has a two--legged
subgraph. But case $(a)$ is impossible: all vertices of $G$ have
even incidence number, so all subgraphs of $G$ have an even
number of external legs.
\herefig{fst3fig13}

Let $E(G)=4$.  
If $E(C_\one)=2$
and $E(C_\two) =6$ $C_\one$ is a two--legged subgraph of $G$. 
If $ E(C_\one)=E(C_\two)=4$, $G$ is as in Figure $\doltwoPI$.
If $C_\one$ or $C_\two$ were 1PR, $G$ would have a two--legged
subgraph, because by Theorem \ONE.2.23, a four--legged 1PR graph
always has a two--legged subgraph.  
\endproof

\chap{Regularity to All Orders}
\noindent
In this chapter, we prove regularity of the counterterm function 
$K$ to all orders in perturbation theory.  In \TWO, we proved
regularity properties of all non--DOL graphs.
We now turn to the problem of larger graphs, where 
the two derivatives applied to $Val(G)$ may act on 
at least two lines. We first show that 
the volume bounds alone are sufficient to control the second derivative
for all two--legged skeleton graphs with $t \geq 2$. 
We then show the same for all graphs with at least three vertices
and $t\ge 1$, except for those already treated in \TWO.
The proof of Theorems \Kalloreg\ and \Salloreg\ 
follows from a combination of these results
with an extension of the methods developed in \ONE.
At the beginning of this Chapter, we recall briefly 
some conventions and results of \ONE\ and \TWO.

\sect{Some Definitions and Results from \ONE\ and \TWO}
We briefly recall the 
equivalence of the theory with the full propagator and the 
initial four-fermion interaction to that with the (physically
relevant) infrared part of the propagator and a bounded and
regular {\it scale zero effective action}, defined in \TWO.2.50,
and give the most basic properties of the single scale propagators.

Integrate out the ultraviolet (that is, large $k_\zer$) part
of the model specified in the Introduction.
By Lemma \TWO.2.3, all vertex functions of the 
resulting scale zero effective
action are $C^{2,\hxp}$, with all derivatives uniformly bounded,
if \AOne{2,\hxp} and \ATwo{2,\hxp} hold. We may therefore assume
that the initial interaction is the scale zero effective action.
Thus our graphs have vertices with even incidence number
(not necessarily four), and vertex functions that are $C^{2,\hxp}$
with bounds uniform in the scales, and the propagators
associated to the lines are the infrared propagators
$$
C_{<0} (p_\zer,E) = \sli_{j<0} C_j (p_\zer,E)
\eqn $$
(see Section \ONE.2.1). 
The single scale propagators $C_j$ are defined in Section \ONE.2.1
(see also \queq{(\TWO.2.53)}) and they satisfy for all $s \le
2$ 
$$
\max\limits_{|\al|=s} \abs{D^\al C_j(p_\zer,e(\p))} \le 
W_s M^{-(s+1)j} \True{\abs{ip_\zer-e(\p)} \le M^j}
\eqn $$
where $D^\al$ is a derivative with respect to $p$ ($\al$ is a multiindex 
with $|\al|=s$, $0\leq s\le k$) of order $s$. 
The indicator functions take the value
$\True{X}=1$ if $X$ is true and $\True{X}=0$ otherwise. 
Here we choose the constant $M$ as in \ONE\ and \TWO, 
namely $M \ge \max\{ 4^3,\sfrac{1}{r_\zer}\}$.

The estimates on scale sums for general labelled graphs, 
will be stated in terms of the function 
$$
\la_n (j,\ep) = \sum\limits_{p=1}^\infty 
(|j|+p+1)^n M^{-\ep p}
.\EQN\LAN $$
It satisfies $(|h|+1)^m \la_n(h,\ep) \le \la_{m+n} (h,\ep)$ and 
$$
\la_m(h,\ep) \la_n(h,\ep) \le \la_{m+n} (h,\ep)
.\EQN\prinq $$
For a proof of 
the second relation, and further properties of $\la_n$, see
Lemma \ONE.2.44.

\ssni
\sect{Outline of the strategy; the string lemma}
\noindent
To make the proof easier to read, we split it into several lemmas
and give an outline of the procedure now.

We do an induction on the order of perturbation theory, with the 
inductive hypothesis that the root scale behaviour of the second 
derivative of the value of any two--legged graph is bounded by a power of 
$|j|$ and that the root scale behaviour of the second derivative
of the localized value of a graph is summable.
All subgraphs, in particular all two--legged proper
subgraphs, of an order $r$
graph are at most of order $r-1$ in the coupling constant $\la$, so the 
induction hypothesis applies to these subgraphs. We show as a first
step that strings of two--legged subdiagrams have the same scale 
behaviour as single propagators.  
 
To every labelled graph $G$, we can associate a skeleton graph $G'$ 
in a natural way by replacing 
strings of two--legged subdiagrams by single lines. By the above, 
the propagators associated to these lines
obey bounds that lead to exactly the same power counting as 
the free propagator. To do the 
induction step, one then has to show that the value of $G'$, as given
by the assignment of propagators and effective vertices, is $C^2$.

We shall first consider the simplified problem in which all vertices are of scale
zero and all lines are of the same scale. We show that all DOL graphs
have double volume gain, so that the root scale behaviour
of the second derivative is still 
summable as $j \to -\infty$. By Theorem \allnonDOL,
this implies the convergence of the second derivative 
for the contribution from all two--legged skeleton graphs except for 
those shown in Figure $\dolone$, $\wicked $, and $\figunin $. 
The multiple sunset graph in Figure $\figunin$ 
is easily treated in the simplified
problem because the scale zero vertices and their derivatives
are all uniformly bounded.  
The values of the graphs 
in Figure $\dolone$ and $\wicked $ were shown to be $C^2$ in \TWO.

After that, we shall put in the full scale structure.
Let $\Ga=(G')\ {\tilde{ }}\, (\ph)$ be the root scale quotient of $G'$ 
(that is, the graph obtained by collapsing all lines that are
not of root scale -- see Definition 2.27 and Remark 2.28 of \ONE). 
If $\Ga $ is DOL, there are two volume gains on root scale, 
which suffices to take two
derivatives and to show H\" older continuity of degree $\ga <1$
of the second derivative. 
Otherwise, $\Ga $ is as in Figure $\dolone$, $\wicked $, 
and $\figunin $. For the graphs as in Figure $\dolone$ or $\wicked $, 
we proceed as in \TWO\ to prevent
the second derivative from acting on root scale. However, it may now act
on the effective vertex function, which now replaces the scale zero 
vertex function as the function $P$ in \queq{\socalled}, and 
whose second derivative does not have a uniform bound in the scales.
Whenever the derivative acts on an effective vertex like this, 
we analyze the graph at the higher scale where the derivative acts 
on a fermion propagator. If the graph is still not DOL at that scale,
it must still be as in Figure $\dolone$ or $\wicked $, and  
we keep going until a second volume gain arises or until we 
end up at scale zero, in which case the results of \TWO\ apply.

The graph in Figure $\figunin$ requires a separate argument which uses that
scale sums over effective vertices with more than four legs grow
and that therefore a gain at higher scale amounts to a gain at 
root scale. This will be discussed in more detail below.
 
We now come to the details. We start with the Lemma about strings. Let 
$$
\abs{g}_{s}^{(j)} = \sum\limits_{\al: |\al| \le s}
\sup\{ \abs{\del^\al g(p)}: \abs{ip_\zer - e(\p)} \le M^j\}
,\eqn $$
and for a differentiable function $T$ defined on the $r_\zer$--neighbourhood
of $S$, let $\sfrac{\del}{\del\th}T$ be defined as 
$$
\frac{\del}{\del\th} T (p) = \nabla T(p) \cdot \del_\th\p(\rh,\th)
\eqn $$
(since $p$ is near to $S$, $p=(p_\zer,\p(\rh,\th)$).

\Lem{\stringprop}{\Lesty Assume \ATwo{2,0}. Let $\veps > 0$, $j<0$, 
$j_\zer, \ldots, j_n \in \{j,j+1\}$
with $\min \{ j_\zer, \ldots , j_n\} =j$, and let
$$
S_j (p) = 
C_{j_\zer}(p_\zer, e(\p)) 
\prod\limits_{k=1}^n T_k^{(j)} (p) C_{j_k}(p_\zer, e(\p))
\EQN\soeinstring $$
Then $\supp S_j \subset \{ p : \abs{ip_\zer - e(\p)} \le M^j\}$. 
If the $T_k^{(j)}$ are $C^2$ and if there are $\ta_k>0$ and 
$n_k \in \N$ such that 
$$
\abs{T_k^{(j)}}_{s}^{(j)} \leq \ta_k \la_{n_k} (j, \veps)
\cases{M^j & $s=0$\cr
1 & $s=1$ \cr
|j|^2 & $s=2$ \cr}
\EQN\bravetes $$
and
$$\eqalign{
\abs{\frac{\del}{\del \th} T_k^{(j)}}_\zer^{(j)} & \le \ta_k \abs{j}^2\;\la_{n_k} (j, \veps)\; M^j
 \cr
\abs{\frac{\del}{\del \th} \frac{\del T_k^{(j)}}{\del p_\al}}_\zer^{(j)} 
& \le \ta_k \abs{j}^2\;\la_{n_k} (j, \veps)
\cr}\EQN\braveretes $$
then 
$$
\sli_{\al:|\al| \le s} \abs{\del^\al S_j(p)} \le (n+1)^s \ M^{-j(1+s)} 
\True{\abs{ip_\zer - e(\p)} \le M^j}
 B_{s,n} \;
\prod\limits_{k=1}^n \ta_k \la_{n_k} (j, \veps)
\EQN\Sgutt $$
with $B_{\zer,n}={W_\zer}^{n+1}$, 
$B_{1,n}= {W_\zer}^n(2W_\zer + W_\one)$ and $B_{2,n}={W_\zer}^{n-1} 
({W_\zer}^2+{W_\one}^2 + W_\zer (2 W_\one+W_\two))$.
The $W_r$ are as in \queq{\shellj}.
Moreover
$$\eqalign{
\abs{\frac{\del}{\del \th} S_j (p_\zer, \p(\rh,\th))} & \le
n {W_\zer}^{n+1} \; |j|^2 \; M^{-j} 
\True{\abs{ip_\zer - \rh} \le M^j}
\prod\limits_{k=1}^n \ta_k \la_{n_k} (j, \veps)
\cr
\abs{\frac{\del}{\del \th} \frac{\del}{\del p_\al}S_j (p_\zer, \p(\rh,\th))} & \le
\tilde W_n |j|^2 \; M^{-2j}   
\True{\abs{ip_\zer - \rh} \le M^j}
\prod\limits_{k=1}^n \ta_k \la_{n_k} (j, \veps)
,\cr}\EQN\thsok $$
with $\tilde W_n = (2n+1)^2 (W_\zer^n W_\one + W_\zer^{n+1})$.
}

\ssni
This Lemma implies that strings of two--legged subdiagrams behave like single
propagators because the bound on the $T_i$ will be proven
for two--legged insertions ($r$-- or $c$--forks or single scale insertions).
\queq{\thsok} means that  derivatives with respect to $\th$ do not affect 
the exponential scaling behaviour, i.e.\ they produce only a factor $|j|$, 
not an $M^{-j}$. For the scale analysis, this is as good as the behaviour 
of the free propagator $C_j$, which satisfies 
$\sfrac{\del}{\del \th} C_j (p_\zer,\p(\rh,\th))=0$.
More precisely, as proven in
Theorem \dreipro, the critical point analysis of \TWO\ is unchanged
under the replacement of $C_j$ by a string $S_j$ satisfying
\queq{\Sgutt} and \queq{\thsok}.

\Proof The support property follows directly from that of $C_j$. Let $s=0$.
By the support properties of $C_j$, 
$$
\abs{S_j}_\zer \le 
\prod\limits_{k=0}^n \abs{C_{j_k}}_{\zer}^{(j)} \; 
\prod\limits_{k=1}^n \abs{T_k^{(j)}}_{\zer}^{(j)}
.\eqn $$
By \queq{\shellj} and \queq{\bravetes}, this is
$$
\le {W_\zer}^{n+1} \; M^{-(n+1)j} \prod\limits_{k=1}^n (\ta_k \la_{n_k} (j, \veps) M^j)
=  {W_\zer}^{n+1} \; M^{-j}\; \prod\limits_{k=1}^n \ta_k \la_{n_k} (j, \veps)
.\eqn $$
Let $s=1$. The derivative in $\sfrac{\del}{\del p_a} S_j$ 
can act on a $C_{j_k}$ or on a $T_k^{(j)}$. The contribution where the 
derivative acts on the product of propagators is bounded by 
$(n+1) {W_\zer}^n W_\one M^{-2j} \prod\limits_{k=1}^n \ta_k \la_{n_k} (j, \veps)$. 
By \queq{\bravetes}, the contribution from the derivative acting on the
product of $T_k^{(j)}$ is bounded by 
$$
n \, {W_\zer}^{n+1} \; M^{-(n+1)j} \left(\prod\limits_{k=1}^n \ta_k \la_{n_k} (j, \veps) \right)
M^{(n-1)j} \le n {W_\zer}^{n+1} \; M^{-2j}\prod\limits_{k=1}^n \ta_k \la_{n_k} (j, \veps) 
.\eqn $$
so
$$
\abs{D^1 S_j}_\zer \le {W_\zer}^n (W_\zer+W_\one) (n+1) 
\; M^{-2j} \; \prod\limits_{k=1}^n \ta_k \la_{n_k} (j, \veps) 
.\eqn $$
Adding $|S_j|_\zer$, we get the desired bound. $s=2$ is similar,
since $M > e$, and therefore $|j| \le M^{-j}$. 
\queq{\thsok} follows from $\sfrac{\del}{\del \th} C_j =0$ and
\queq{\bravetes}. \endproof

\sect{Volume estimates for DOL graphs}
\noindent
We continue with the volume improvement bounds
for \OL\ and doubly overlapping graphs. 
\def\gOL{{g_{OL}}}
Recall that $r_\zer>0$ determines the size of the neighbourhood of
the Fermi surface where radial and angular variables are 
introduced (see Section 2.2 of \TWO). Without loss of generality,
we may assume $r_\zer <1$.

\Lem{\DOLVol}{ \Lesty Assume \ATwo{2,0}--\AFou. 
Let $\veps\in (0,r_\zer)$ with $|\log \veps| >1$ and $G$ be a
skeleton graph with oriented lines. Let 
$$
\gOL(G)=\cases{ 0 & if $G$ is \NOL\cr
                  1 & if $G$ is \OL\ but not DOL\cr
                  2 & if $G$ is DOL\cr}
\EQN\gOLdef $$
and let $T$ be a spanning tree of $G$; if $G$ is DOL let $T$ give rise
to two separate overlaps.
To every external leg $\ell$ of $G$ we associate
a momentum $\r_\ell\in\cB$. To every line $\ell\in L(G)\setminus L(T)$
we associate an integration variable $\th_\ell$ running over the 
unit sphere $S^{d-1}$ and 
a momentum $\p_\ell=\p(0,\th_\ell)$. To every line $b \in L(T)$ we 
associate a momentum $\q_b$ in the usual way, 
i.e.\ we choose an endpoint of $b$
(it does not matter which one), and set $\q_b=\p_{b,in}-\p_{b,out}$, 
where $\p_{b,in}$ is the sum of $\p_\ell$ and $\r_{\ell'}$ from the 
incoming lines except $b$, and $\p_{b,out}$ is the sum of $\p_\ell$ 
and $\r_{\ell'}$ from the outgoing lines except $b$.  Then 
the integration volume 
$$
\cV=\sup\limits_{\r_\ell} \;
\int\prod\limits_{\ell\in L(G)\setminus L(T)} 
\frac{d\th_\ell}{|S^{d-1}|}
\prod\limits_{b\in L(T)} \True{|e(\q_b)| \le \veps }
\EQN\fourfour $$
satisfies
$$
\cV \le \cW(\veps)^{\gOL(G)}
\le \cases{  \left( Q_V \;\veps |\log\veps |\right)^{\gOL(G)} & $d=2$ \cr
 \left( Q_V\;\veps\right)^{\gOL(G)} & $d \ge 3$}
,\EQN\fourfive $$
and
$$\eqalign{
\tilde \cV &= \sup\limits_{\r_\ell} \;
\int\prod\limits_{\ell\in L(G)\setminus L(T)} 
\left( d\p_\ell \; \True{|e(\p_\ell)|\le \veps}\right)
\prod\limits_{b\in L(T)} \True{|e(\q_b)|\le \veps} \cr
& \le
(\nu_\zer \veps)^{|L(G)|-|L(T)|} \; 
\cases{
(\nu_\one\veps |\log \veps|)^{\gOL(G)} & $d=2$ \cr
(\nu_\one\veps )^{\gOL(G)} & $d\ge 3$ \cr}
\cr}\EQN\foursix $$ 
with
$$\eqalign{
\nu_\zer &= 2 |S^{d-1}| |J|_\zer\cr
\nu_\one &= 2 Q_V (1+2\sfrac{|e|_\one}{u_\zer})^2
.\cr}\eqn $$
Here $J$ is the Jacobian of the change of variables to $(\rh,\th)$ and
$Q_V$ is the constant in Theorem $\bestvol$.}

\Proof The proof is a simple extension of that of Lemma \ImpPow.
If $G$ is \NOL, the right hand side of \queq{\fourfive} is one. The bound 
is obviously obtained by dropping the product over $b\in L(T)$
in \queq{\fourfour}.
If $G$ is \OL, but not DOL, let $\ell_\one$ and $\ell_\two$ be two 
loop lines whose loops overlap on $b_\zer \in L(T)$. Then 
$\q_{b_\zer}=v_\one\p(0,\th_{\ell_\one})+ v_\two\p(0,\th_{\ell_\two})+\Q$
with $v_i \in \{1,-1\}$ and $\Q$ possibly depending on the $\th_\ell$,
$\ell$ different from $\ell_\one$ and $\ell_\two$, and on the external 
momenta. So 
$$\eqalign{
\cV &\le \sup\limits_\r \int\prod\limits_{\ell\not\in\{\ell_\one,\ell_\two\}}
\frac{d\th_\ell}{|S^{d-1}|} 
\int d\th_{\ell_\one}\; d\th_{\ell_\two} \True{|e(\q_{b_\zer})|\le \veps} \cr
&\le \sup\limits_\r \sup\limits_{\th_\ell,\ell\not\in\{\ell_\one,\ell_\two\}}
\int d\th_{\ell_\one}\; d\th_{\ell_\two} \True{|e(\q_{b_\zer})|\le \veps} 
.\cr}\eqn $$
The last integral is bounded, uniformly in $\Q$, by $\cW(\veps)$, so
$\cV \le \cW(\veps)$, and \queq{\fourfive} follows from Theorem \bestvol.

If $G$ is DOL, let $k_\one,k_\two,\ell_\one,\ell_\two$ and $a,b\in L(T)$
such that the loops of $k_\one$ and $k_\two$ both contain $a$, 
and such that the loops of $\ell_\one$ and $\ell_\two$ both contain $b$
but not $a$. We use 
$$
\prod\limits_{l\in L(T)} \True{|e(\q_l)|\le \veps} \le
\True{|e(\q_a)|\le \veps} \; \True{|e(\q_b)|\le \veps} 
\EQN\mmm $$
and proceed as in the previous case, to bound 
$$
\cV \le \sup\limits_\r\; 
\sup\limits_{\th_\ell,\ell\not\in\{k_\one,k_\two,\ell_\one,\ell_\two\}}
S_{ab}(\veps)
\eqn $$
with 
$$
S_{ab}(\veps) = \int d\th_{k_\one}\; d\th_{k_\two}\; d\th_{\ell_\one}\;
d\th_{\ell_\two}\; \True{|e(\q_a)|\le \veps} \; \True{|e(\q_b)|\le \veps} 
\eqn $$
and 
$$\eqalign{
\q_a &= v_\one \p(0,\th_{k_\one})+v_\two \p(0,\th_{k_\two})+\Q_a \cr
\q_b &= v_\one \p(0,\th_{\ell_\one})+v_\two \p(0,\th_{\ell_\two})+\Q_b 
.\cr}\eqn $$
As before, $\Q_a$ and $\Q_b$ may depend on many other momenta.
However, $\Q_a$ does not depend on $\th_{\ell_\one}$ or
$\th_{\ell_\two}$ because the overlaps are separate. Thus 
$$\eqalign{
S_{ab}(\veps) &= \int d\th_{k_\one}\; d\th_{k_\two}\;
\True{|e(\q_a)|\le \veps} \;
\int  d\th_{\ell_\one}\;d\th_{\ell_\two}\;  \True{|e(\q_b)|\le \veps} \cr
&\le \int d\th_{k_\one}\; d\th_{k_\two}\;
\True{|e(\q_a)|\le \veps} \; \cW(\veps) \le 
\cW(\veps)^2
\cr}\EQN\oho $$
which proves the first inequality of \queq{\fourfive}. The second 
inequality follows by Theorem \bestvol.

The proof of \queq{\foursix} is trivial in the \NOL\ case and similar 
in the two others; we do the DOL case. We first change variables to 
$(\rh_\ell,\th_\ell)$
for all $\ell \in L(G)\setminus L(T)$. This is possible because
$\abs{\veps_l} < r_\zer$ for all $l$, so the support of every
loop integral is contained in the neighbourhood of $S$ where
$\rh$ and $\th$ are defined. Using \queq{\mmm}, 
$$
\tilde \cV \le |J|_\zer^{|L(G)|-|L(T)|} \;
\int \prod\limits_{\ell\in L(G)\setminus L(T)} d\rh_\ell
\; \True{|\rh_\ell|\le \veps } \; 
\prod\limits_{\ell\in L(G)\setminus L(T)} d\th_\ell\; 
\True{|e(\q_a)|\le \veps} \; \True{|e(\q_b)|\le \veps}
.\eqn $$
Now $\q_a$ depends also on $\rh_{k_\one}$ and $\rh_{k_\two}$, and 
$\q_b$ depends also on $\rh_{\ell_\one} $ and $\rh_{\ell_\two} $,
and possibly also on $\rh_{k_\one}$ and $\rh_{k_\two}$. We use $|\rh_\ell|
\le \veps$ and Taylor expansion to get 
$$\eqalign{
\True{|e(\q_a)|\le\veps} & = \True{|e(\pm\p(\rh_{k_\one},\th_{k_\one})
\pm \p(\rh_{k_\two},\th_{k_\two}) +\Q_a)|\le \veps} \cr
&\le \True{|e(\pm\p(0,\th_{k_\one})
\pm \p(0,\th_{k_\two}) +\Q_a)|\le \veps (1+2\sfrac{|e|_\one}{u_\zer})}
\cr}\EQN\fourfift $$
and similarly, 
$$
\True{|e(\q_b)|\le\veps} \le \True{|e(\pm\p(0,\th_{k_\one})
\pm \p(0,\th_{k_\two}) +\Q_b)|\le \veps (1+2\sfrac{|e|_\one}{u_\zer})}
.\EQN\foursixt $$
We integrate over $\th_{k_\one},\th_{k_\two},\th_{\ell_\one},\th_{\ell_\one}$
first, to get a subintegral similar to $S_{ab}$, and take a sup over all 
other $\th$--variables. Using \queq{\oho} and Theorem \bestvol, and collecting
the constants, we get \queq{\foursix}. \endproof

\noindent
We state a generalization of this Lemma where the support condition may 
depend on the line, i.e.\ $\veps$ gets replaced by $\veps_\ell$ on 
every line. 
\Lem{\DOLVolnoamol}{\Lesty  Assume \ATwo{2,0}--\AFou.
Let $G$ be a skeleton graph with oriented lines
and let $T$ be a spanning tree of $G$; if $G$ is DOL let $T$ give rise
to two separate overlaps. Let the momentum assignments be as in Lemma 
$\DOLVol$, associate an $\veps_\ell\in (0,r_\zer)$ to every line $\ell\in L(G)$,
and let $\underline{\veps}=\min\limits_{\ell\in L(G)} \veps_\ell$.
Then 
$$
\bar \cV (G,T) = \int\prod\limits_{\ell\in L(G)\setminus L(T)} 
\left( d\p_\ell \; \True{|e(\p_\ell)|\le \veps_\ell}\right)
\prod\limits_{\be\in L(T)} \True{|e(\q_\be)|\le \veps_\be}
\eqn $$
satisfies
$$
\bar \cV (G,T)\le \left(\nu_\one |\log\underline{\veps}|\right)^{\gOL (G)} \; 
\left(\prod\limits_{\ell\in L(G)\setminus L(T)} 
\nu_\zer \veps_\ell\right) \;\;\cases{
1 & if $G$ is \NOL\cr
\veps_a & if $G$ is \OL, but not DOL \cr
\veps_a\; \veps_b & if $G$ is DOL.}
\eqn $$
$a$ and $b$ are lines of $T$ common to the two (separate) pairs of
overlapping loops.
For $d \ge 3$ the factor $|\log\underline{\veps}|^{\gOL(G)}$ is absent.
}
\Proof The only difference to the previous proof is that 
 one has to be careful in the 
Taylor expansion steps \queq{\fourfift} and \queq{\foursixt}
because $\veps_l$ now depends on $l$. We only 
discuss the modifications because of this. The point is that it may happen
that the line in the tree is on a lower scale than those in the loops,  
i.e.\ that in the integral 
$$
\cV_\two = \int d\p_\one\; d\p_\two \; \True{|e(\p_\one)|\le \veps_\one}
\True{|e(\p_\two)|\le \veps_\two}\; 
\True{|e(v_\one\p_\one+v_\two\p_\two+\Q)| \le \veps_\thr}
\eqn $$
$\veps_\thr < \veps_\one$, or $\veps_\thr < \veps_\two$, in which case
the Taylor expansion does not give \queq{\fourfift}. Assume, without loss
of generality, that $\veps_\two = \max \{ \veps_\one,\veps_\two,\veps_\thr\}$.
Change variables from $\p_\two$ to $\p_\thr=v_\one\p_\one+v_\two\p_\two+\Q$,
which is then also integrated over $\cB$, so that
$$
\cV_\two = \int d\p_\one\; d\p_\thr  \True{|e(\p_\one)|\le \veps_\one}
\True{|e(\p_\thr)|\le \veps_\thr} 
\True{|e(v_\two\p_\thr-v_\one v_\two \p_\one- v_\two \Q)| 
\le \veps_\two}
.\eqn $$
Changing to variables $\rh$ and $\th$, the Taylor expansion argument 
in $\rh_\one$ and $\rh_\thr$ now works since 
$\veps_\two \ge \max \{ \veps_\one,\veps_\thr\}$. Therefore
$$
\cV_\two \le (2|J|_\zer)^2 \veps_\one\;\veps_\thr 
\cW\left( (1+2\sfrac{|e|_\one}{u_\zer})\veps_\two\right)
\le (2|J|_\zer)^2 \veps_\one\;\veps_\thr \; 
\nu_\one \veps_\two \; |\log \veps_\two|
\eqn $$
The factor $\log\veps_\two$ is bounded by $|\log \underline{\veps}|$.
In $d\ge 3$ this factor is absent. 
Note that in the DOL case, the momentum $\Q_a$ is independent of the 
integration momenta $\p_\one=\p_{\ell_\one}$ and $\p_\two=\p_{\ell_\two}$
of the inner loop and so is not affected by the change of variables 
to $\p_\thr$. \endproof

\sni
In the following, we assume that $r_\zer $ is chosen so small that all 
bounds from \TWO\ and the volume bounds apply, and such that
$$\eqalign{
|\log r_\zer | & \ge 1 \cr
\nu_\one r_\zer |\log r_\zer | \log M < 1
.\cr}\EQN\arziro $$

\sect{Doubly volume--improved power counting for skeleton graphs}
Let $m\ge 1$ and $G$ be a skeleton graph with 
$E(G)=2m$ external legs and $n$ vertices $v_\one,\ldots ,v_n$. 
Denote the incidence number of vertex $v$ by $2m_v$ (where $m_v\ge 2$).
To every vertex we associate a $C^2$ vertex function 
$$
\cU_v: (\R\times \cB)^{2m_v-1}\times \{ \uparrow, \downarrow\}^{2m_v} \to \C
\eqn $$
that is totally antisymmetric under simultaneous exchange of momenta and spins
(see Definition 2.10 $(ii)$ and Definition 2.8 $(ii)$ and $(iii)$ of \ONE)
and define the norm
$$
\abs{\cU_v}_s = \sum\limits_{\al:|\al|\le s}
\sup\limits_{p_\one,\ldots p_{2m_v-1}} \; 
\max\limits_{A \in  \{ \uparrow, \downarrow\}^{2m_v}}
\abs{D^\al \cU_v (p_\one,\ldots,p_{2m-1},A)}
.\eqn $$
We start with an estimate where all lines of the graph have the
same scale. 

\Lem{\ssBound}{\Lesty  Assume \ATwo{2,0}--\AFou.
Let $j<0$ and $G$ be a skeleton graph, 
with $\gOL$ defined as in \queq{\gOLdef}. 
To every line $\ell$, associate a propagator $S_\ell (p_\ell) \; 
\de_{\al\al'}$, and define the value of $G$ in the standard way 
(see Definition $2.10$ $(ii)$ of \ONE). 
If for all $\ell \in L(G)$ and $s\le 2$ there are constants
$P_{\ell,s} \le 1$ such that for any $\si=(\si_\zer,\ldots \si_d)$
with $|\si|=\si_\zer+\ldots+\si_d=s$, 
$$
\abs{D^\si S_\ell (p)} \le P_{\ell,s} M^{-j(1+s)}
\True{\abs{ip_\zer-e(\p)}\in \lbrack M^{j-2},M^j\rbrack }
,\EQN\Prophyp $$
then
$$
\abs{Val(G)}_s \le  2 (4\nu_\zer)^{|L(G)|} \; 
 M^{j(2-m-s)}\;  \left(\nu_\one  \; |j|M^j\; \log M\right)^{\gOL(G)} \; 
\mu_s(G)
\EQN\ssbound $$
with 
$$
\mu_s(G)= \max\limits_T \sum\limits_{\si\in Z(s,T,G)}  
\prod\limits_{l \in L(G)} P_{l,|\si_l|}\; 
\prod\limits_{v\in V(G)} \left( \abs{\cU_v}_{|\si_v|} \, 
M^{j(m_v-2+|\si_v|)}\right)
\EQN\ssconst $$
The maximum in \queq{\ssbound} is over all spanning trees $T$ of G,
and $Z(s,T,G)$ is the set of multiindices 
$$
\si:\; (L(G) \times\natz{d}) \times
\prod\limits_{v\in V(G)} \left( \nat{2m_v} \times \natz{d} \right) 
\quad \to \{ 0,1,2\}
\eqn $$
such that 
$$
\sum\limits_{\ell \in L(G)} |\si_\ell| +  \sum\limits_{v\in V(G)}
|\si_v| \le s
,\eqn $$
$\si_\ell = 0$ if $\ell \not\in L(T)$, and $\si_{v,k,r}=0$ if the 
leg $k$ of $v$ is not hooked to a 
line in $L(T)$. Here $|\si_\ell| = \sum\limits_{i=0}^d \si_{\ell,i} $
and $|\si_v|= \sum\limits_{v,k,i} \si_{v,k,i}$.
The constants are as in Lemma $\DOLVol$.}

\Rem{\dummy} Up to powers of $|j|$, $M^{-j(m_v+\si_v -2)}$
is precisely the standard power counting behaviour of a $\cU_v$  
given by an effective vertex of the model on scale $j$, proven in 
Theorem 2.45 of \ONE\ for $\si_v \le 1$ for arbitrary (not
necessarily skeleton) graphs. 
Inspection of the proof of that theorem shows that for skeleton graphs, 
it also holds for $\si_v =2$.  We shall show below that this scaling 
behaviour holds for $\si_v=2$ and arbitrary graphs, so there will 
be no need to go through the proof of Theorem 2.45 in \ONE. 

\Proof Let $T$ be a spanning tree for $G$. If $G$ is DOL then let $T$
give rise to two separate overlaps. 
Fixing the momenta on the lines of $T$, we obtain (denoting all spin 
indices by $A$ and $A_v=(\al_\one^{(v)},\ldots,\al_{2m_v}^{(v)})$) 
for a multiindex $\be $ with $|\be|\le 2$
$$\eqalign{
D^\be Val(G^J) (q_\one,\ldots,q_{2m-1},A) & = 
\sum_{\si}
m(\be,\si) \sum _{A_\one,\ldots,A_v} \;
\int \prod\limits  _{l \in L(G)\setminus  L(T)} \db^{d+1}p_{l} 
\; \prod\limits_{l \in L(G)}
D^{\si_l} S_l (p_{l}) \de_{\al_{l} \al_{l}^{\prime}} \cr
& \prod\limits_{v\in V(G)} \;  D^{\si_v} \cU _{v} 
(p_\one^{(v)}, \ldots , p_{2m_v-1}^{(v)},A_v) 
\cr} \EQN\DbeValG $$
where the sum is over multiindices $\si$ and the possible values of 
the assignment of $\si_{l,r}$ and $\si_{v,r}$ depend on the multiindex $\be$,
but the restrictions stated in the Lemma apply because only the momenta
on lines in the tree can depend on the external momenta. The spin indices 
$\al_l$ and $\al'_l$ are fixed by the corresponding indices of the 
vertex functions at the endpoints of the line.
$m(\be,\si )$ is a multinomial factor, and for each $t \in L(T)$, 
$p_t$ is a linear combination 
of the loop momenta $(p_l)_{l \in L(G)\setminus L(T)}$ and, possibly,
of the external momentum $q$.

The Kronecker delta for the spins in the propagator implies that there 
is only one spin sum per line. We bound this sum by 2 times the supremum
over all spin values, and use \queq{\Prophyp} to bound the propagators.
We also use that $m(\be,\si) \le 2$. This gives
$$
\abs{D^\be Val(G^J) }_\zer \le 2 \cV(G,T) \; 2^{|L(G)|} Y(\be,T)
\EQN\fthn $$
with
$$
Y(\be,T) = \sum_{\si}
\prod\limits_{l\in L(G)} P_{l,|\si_{l,r}|} M^{-j(1+|\si_{l,r}|)}\;
\prod\limits_{v\in V(G)} \abs{\cU_v}_{|\si_{v,r}|}
\EQN\ffty $$
and the integration volume 
$$
\cV(G,T) = \int\prod\limits_{l\in L(G)\setminus L(T)} \db^{d+1} p_l\;
\prod\limits_{l\in L(G)} \True{|i(p_l)_\zer-e(\p_l)|\le M^j}
.\eqn $$
Collecting the exponent of $M^j$ and noting that 
$\sum_{v,r} \si_{v,r}+ \sum_{l,r} \si_{l,r}= |\be| $, we get
$$
Y(\be,T) =M^{-j(|L(G)|+|\be|)}\;
M^{j\sum_v(2-m_v)}\;  \sum_{\si}  
\prod\limits_{l\in L(G)} P_{l,|\si_l|}
\prod\limits_{v\in V(G)} \left(\abs{\cU_v}_{|\si_v|}\, 
M^{j(m_v-2+|\si_v|)}\right)
.\eqn $$
We  sum over all $\be$ with $|\be|\le s$ so that the sum over $\si $ now 
runs over the set $Z(s,T,G)$, and bound $M^{-j|\be|}\le M^{-js}$.
Taking the maximum over all spanning trees $T$ of $G$, we get
$$
Y(\be,T) \le M^{-j(|L(G)|+s)}\;
M^{j\sum_v(2-m_v)}\; \mu_s(G)
.\eqn $$
After all this combinatorics, we turn to the essential part of the estimate,
the bound for $\cV(G,T)$. Because $|ip_\zer -e(\p)| \le M^j$ implies
$|p_\zer|\le M^j$ and $|e(\p)|\le M^j$, 
$$
\cV(G,T) \le  \; \tilde \cV \; \int\prod\limits_{l\in L(G)\setminus L(T)}
d(p_l)_\zer \; \True{|(p_l)_\zer|\le M^j}
\eqn $$
with $\tilde \cV$ given in \queq{\foursix}. By \queq{\foursix},
$$
\cV(G,T) \le \left(2 \nu_\zer M^{2j}\right)^{|L(G)|-|L(T)|} \; 
(\nu_\one \; |j| M^j\; \log M )^{\gOL(G)}
.\eqn $$
Rearranging the various factors, 
using that $|L(G)|= \sum_v m_v -m$ and 
$|L(T)|=|V(T)|-1=\sum_v 1 \; -1$, we obtain the result. 
\endproof

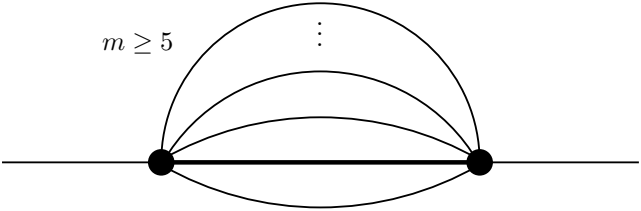
\sect{The multiple sunset}

\noindent
Before considering the full scale structure, we show how the estimates
for the multiple sunset
graph shown in Figure $\figunin$ fit into the picture. The point is here
that because there are five or more lines connecting the two vertices, 
the strategy from the second--order case cannot be used directly, 
because there are now at least five fermion momenta whose sum is to be near
to the Fermi surface instead of three. 
Also, on root scale, there is only one volume gain from any of the two--loop
subintegrals that one can choose, and one volume gain alone is not sufficient 
to cancel the large factors arising from two derivatives acting
on root scale. However, the values of these graphs are $C^2$ again 
by volume effects only, by the following argument. 
The scale zero vertices of the model have at most four legs, 
so the vertex functions $\cU_v$ associated to the two effective
vertices on root scale
must be values of subdiagrams with $2m \ge 6$ external legs. 
There are two possibilites:

\leftit{$(i)$} both effective vertices are scale zero effective vertices

\leftit{$(ii)$} at least one of them is not a scale zero effective
vertex, and thus corresponds to a subgraph with lines carrying
scales $j_l \le -1$. 

\ssni
In case $(i)$, the value of $G$ is (up to an unimportant sign
factor) 
$$
Val(G) = \int \left(\pli_{l=2}^i \db p_l\; S_{l,j}(p_l)  \right)
S_{1,j} (q+\sli_{l=2}^i v_l p_l) \; 
\cU_\one(q,\underline{p}) \cU_\two (q,\underline{p})
\EQN\AroundEq $$
where $i$ is the number of lines joining vertex 1 to vertex 2,
$\cU_k$ is the vertex function associated to vertex $k$, 
$v_l \in \{1,-1\}$, and the $S_{l,j}$ satisfy \queq{\Prophyp}. 
Since $G$ is a {\it multiple} sunset, $i \ge 5$. 
The derivatives of $\cU_\one$ and $\cU_\two$ are uniformly
bounded, so the worst case is when all derivatives act on $S_{1,j}$.
By \queq{\Prophyp}, 
$\max\limits_{|\al|=s} \abs{D^\al S_{1,j}}_\zer \le P_{1,s} M^{-(1+s)j}$,
so, using this estimate after taking at most two derivatives and 
estimating the integrals in the standard way (not even using
any extra volume effect), we have 
$$
\abs{Val(G)}_\two \le \Const M^{-3j} M^{(i-1)j} \le 
\Const M^{-3j} M^{4j} \le \Const M^j
\eqn $$  
with the constant given in terms of the $P_{l,s}$ with $s \le 2$.
Thus the sum over $j<0$ is convergent. Moreover, the procedure of 
taking differences of Section \TWO.3.4 implies H\" older continuity
of degree $\hxp$ if \AOne{2,\hxp} and \ATwo{2,\hxp} hold. 
In brief, since the scale zero effective vertex functions with
$2m \ge 6$ external legs are bounded, i.e. behave as $1=M^{0(2-m)}$
instead of growing like $M^{j(2-m)}$, the number of integrations
alone already suffices to make two derivatives converge. 

\sni
Case $(ii)$: We shall show in the next lemma that the graph must be DOL at 
a higher scale $h < 0$. In general, a gain at a scale $h$ does not 
result in a gain at a lower scale $j$, once $h$ is summed down to $j$.
But the subgraphs here have $m \ge 3$, and thus the root scale behaviour
of their values is without improvement $M^{h(2-m)} \ge M^{-h}$, 
i.e. it grows because $m \ge 3$. {\it Therefore, a gain at the higher scale $h$
slows down the rate of growth of the scale sum that gives $\cU_v$}. 
Using
$$
\sum_{h\ge j} M^{h(2-m)} \; M^{\ep h} \le 
\Const (\ep,M) M^{j(2-m)}  \; M^{\ep j} 
\quad \quad \hbox{ for all }\quad  2+\ep-m < 0
\EQN\DownSum $$
it is `transported' down 
to a gain at scale $j$ when $h$ is summed down to $j$. 
The dominant term in $\sum\limits_{h\ge j} M^{h(2-m)} M^{\ep h}$
is that with $h=j$. On the other hand, 
for $m=1$ the scale sum is 
already convergent, so that only the speed of convergence, but not 
the scale behaviour of the sum, is changed by the improvement factor
$M^{\ep h}$.
For $m=2$, the same holds since the improvement factor $M^{\ep h}$
only removes a polynomial growth in $\abs{j}$. 
Thus, for $m \le 2$, the dominant term in \queq{\DownSum} is that with $h=0$.
In the following Lemma, we prove that for the graphs of Figure $\figunin$,
the gain arising at a higher scale can indeed be transported to root scale.

To formulate the lemmas that follow, we now need the definitions of 
the tree formalism  (see \quref{FT1,2} or \ONE). We also assume 
familiarity with the results of Sections 2.4 -- 2.6 of \ONE, although
we shall explain the most essential notions briefly.

Let $\cT$ be a tree. The vertices of $\cT$ with incidence number $1$
are called leaves; the others are the forks of $\cT$.
Let $\cT$ be rooted at a fork $\ph$ and have $n$ leaves, 
and let $\pi$ be the predecessor map
that maps every vertex $f\ne\ph$ to the unique fork $\pi(f)$ of $T$
whose distance from $\ph$ (measured in steps over lines of $T$) 
is one less than that of $f$. $\cT$ is compatible to $G$, 
$\cT \sim G$, if there is a family $(G_f)_{f\in \cT}$ of connected subgraphs 
of $G$ such that 
\leftit{} $G_\ph=G$, 
\leftit{} for all $f$ and $f'$, either $G_f \subset G_{f'}$ or
$G_{f'} \subset G_f$ or $G_f \cap G_{f'} = \emptyset$. 
\leftit{} $G_f$ is a subgraph of $G_{f'}$ if and only if 
$f'$ is between $\ph$ and $f$ in the ordering of the tree.
\leftit{} if $\pi(f)=f'$ then $G_f$ is a proper subgraph of $G_{f'}$

\ssni
Let $j<0$ and define
$$\eqalign{
\cJ(\cT,j) = \{ (j_f)_{f\in \cT} : 
j_\ph=j, \hbox{ and for all } f\in \cT\setminus \{ \ph\}
\quad j_f\in\{j_{\pi(f)}+1,\ldots,-1\}\}
\cr}\eqn $$
This set is well--defined because of the recursive structure of $\cT$ 
given by $\pi$. Note that scales are also associated to the the leaves 
of $\cT$; we shall need this because these leaves may be forks of 
the original Gallavotti-Nicol\` o trees of our model, at which the latter
have been trimmed. Every element of $\cJ(\cT,j)$ defines a labelling 
$J: L(G) \to \{ j, \ldots, -1\}$, $\ell \mapsto j_\ell$, where 
$j_\ell = j_f$ for all $\ell$ that are in $L(G_f)$ but not in $L(G_{f'})$
for any $f'$ with $\pi(f')=f$. Equivalently, 
$$
j_\ell = \max\{ j_f : \ell \in L(G_f)\}
.\eqn $$
Given an assignment of propagators to 
the lines of $G$ and vertex functions to the vertices of $G$,
the value of the labelled graph $G^J$ 
is defined in the standard way (see Definition 2.10 $(ii)$ of \ONE\
and the discussion in Lemma \ssBound).

For a subtree $\cT'$ of $\cT$, the quotient graph $\tilde G (\cT')$ 
is the quotient graph of $G_{\ph_{\cT'}}$ in which, for all leaves 
$b$ of $\cT'$, the subgraph $G_b$ of $G$ is replaced by a vertex with 
the same incidence number (see Section 2.6 of \ONE\ for details).

\Lem{\Fnine}{\Lesty Let $G$ be a skeleton graph with at least 
three vertices, $j<0$, $\cT$ be a tree,
and $J$ a labelling of $G$ consistent with $\cT$ and such that the 
root scale is $j$. Assume that $\tilde G(\ph)$ is a multiple sunset
(see Figure $\figunin $). 
Let the forks of $\cT$ corresponding to the two vertices 
$v_\one$ and $v_\two$ in Figure $\figunin $ be $f_\one$ and $f_\two$, 
and assume that $G_{f_\one}$ contains an additional vertex of $G$. 
There exists a subtree $\cT'$ of $\cT$, consisting of forks 
$\ph < f_\one < f_\thr < \ldots < f_n < f'$ such that for all 
$k \in \{ 3,\ldots,n\}$, $f_k$ has incidence number $2$ on $\cT$, 
and $f'$ has incidence number at least $3$ on $\cT$. Moreover, 
$E(G_{f_\one}) = 2 m_{f_\one} \ge 6$, and 
$E(G_{f_k}) = 2m_{f_k} \ge 2 m_{f_{k-1}}+2$ for all $k \ge 3$. 
The spanning tree for $\tilde G(\cT')$ can be chosen such that one 
overlap takes place on scale $j$.  The other gain arises on scale
$j'$, where $j'=j_{f'}$. $j'$ is also the lowest scale on the 
tree $\cT'$ where a derivative
with respect to the external momentum can act on any line in $G_{f_\one}$.}

\ssni
Note that the statement is not that the subgraph $G_{f_\one}$ 
itself becomes overlapping at scale $j'$. It is the entire graph 
that becomes DOL at that scale. See Figure \nextfig{\figufrt} $(c)$
for an example how the graph $\tilde G (\cT_{f_\one}')$ associated
to the effective vertex $v_\one$ may look.

In Lemma \Fnine, we do not assume that the vertices of $G$ are four--legged.
If they are, then $G$ will always have at least four scale zero vertices if 
$\tilde G(\ph)$ is of the form of Figure $\figunin$.

\Proof If $G$ were not DOL at any scale, then, by Theorem \allnonDOL,
it would have to be a \WL\ (see Figure $\wicked $), 
since it has at least three vertices. But no quotient graph of a \WL\
can be a multiple sunset,
so $G$ has to be DOL at some scale, and there a minimal scale 
$h \le 0$ on which $G$ is DOL. We now show the  
more detailed statements. As before, we denote by $t_G \ge 0$ the 
number of steps required to walk from one external vertex to the 
other over lines of $G$. By assumption, $t_{\tilde G(\ph)} = 1$, 
so any quotient graph $\bar G$ of $G$ containing the lines of 
$\tilde G(\ph)$ has $t_{\bar G} \ge 1$.
We grow the tree $\cT'$ by adding forks and leaves to $\ph$ 
by the following algorithm (the various possibilities that can
arise during the procedure are sketched in Figure $\figufrt$).
In the first step, we add the fork $f_\two$ 
to get $\cT' = \matrix{f_\two\cr | \cr \ph\cr}$
and consider the graph $G'=\tilde G(\cT')$. 
If $G'$ has more than two vertices, $\cT'$ is complete.

\herefig{fst3fig14}

If $G'$ still has only two vertices, then going from $\tilde G(\ph)$ 
to $G'$ has only uncovered some self--contractions of a vertex
$v_\two'$ that replaces $v_\two $ in $G'$ (see Figure $\figufrt$ $(a)$). 
If a graph $H_\one$ 
is obtained from $H_\two$  by a self--contraction (tadpole line),
$H_\two$ must have incidence number at least two more than $H_\one$. 
Therefore the fork $f_\two'$ with $\pi(f_\two')=f_\two$ must have
$E(G_{f_\two'}) \ge 8$ (there is only one such fork, 
otherwise there would have been 
more than one effective vertex in $G_{f_\one}$). 
In this case we go on by adding $f_\two' $ to 
$\cT$ as a leaf above $f_\two$. If the number of 
vertices still does not increase, we repeat this procedure, 
and we stop when a new vertex appears, so that $G'$ has more than
two vertices. Then $\cT'$ is complete. 
By construction, all forks $f>\ph$ of $\cT'$ have $E(G_f) \ge 6$.
If at no scale a new vertex appears, then the subgraph $G_{f_2}$ consists 
only of a single vertex with self--contractions as lines. In that case,
we start from the beginning by resolving $G_{f_\one}$, i.e.\
we apply the same procedure to $\matrix{f_\one\cr | \cr \ph\cr}$.
Then a third vertex must appear at some scale because $G$ has at least
three vertices. 

By the symmetry of the graph, we may assume without loss 
that the above procedure has produced a third vertex, and 
thus a suitable candidate $\cT'$, by resolving $G_{f_\two}$.

We split the proof that $G'$ and $\cT'$ have the desired properties
into two cases. Recall that under our conventions, a path in $G$
is a non-selfintersecting walk over the lines of $G$, i.e.\ no vertex is 
visited twice by a path (see Definition 2.17 of \ONE).
$G'$ consists of $v_\one$ and a subgraph 
$G_\two'=\tilde G_{f_\two}(f_\two)$, i.e.\ when $G_\two'$ is 
collapsed to a point, $G'$ becomes $G$.

(1) There is no path in $G'$ between the external vertices
that contains at least two lines.  This implies that all the extra 
vertices of $G'$ can only be connected to $v_\two'$, so that $G_\two'$
takes the form shown in Figure $\figufrt$ $(b)$ (not all  
the subgraphs drawn in that figure 
have to be there, but at least one of them 
must be there). All connected components $C_\al$ of $G'_\two - v_\two'$ 
must have at least four external legs because all quotient 
graphs of $G$ are skeleton graphs. This is also the reason why 
generalized self--contractions cannot appear. It is obvious that 
the subgraph $G_\two'$ is \OL, hence 
$G$ is DOL. Choosing any spanning tree for the subgraphs and combining
them with the fat lines in Figures $\figunin$ and $\figufrt$ $(b)$, 
the statement of the Lemma is obvious.

(2)  There is a path in $G'$ between the  external vertices 
with at least two lines (even though the {\it shortest} path 
may still be of length one). If $t_{G'} \ge 2$, then $G'$ is as 
shown in figure $\figusev$ or 8, so it is DOL. Taking the quotient graph of 
$G'$ to get back $\tilde G(\ph)$, we see that one of the lines 
carrying the overlaps must be at scale $j$. That leaves $t_{G'}=1$, i.e.\ the 
shortest path still has length one. By construction of $G'$, one of 
the lines of $\tilde G(\ph)$, $\ell_\one$, is the first step in a walk 
of length at least two from $v_\one$ to the other external vertex, 
$v'_\two$, of $G'$. 
$\ell_\one$ carries scale $j_{\ell_\one}=j$. Call $w$ the other
endpoint of $\ell_\one$. Collect all other vertices of $G_\two$ into a graph 
$D_\two$, and let $\Ga_\two$ be the connected component of $D_\two$ 
that contains $v_\two'$. Collect all other connected components together
with $w$ and the lines between them into a connected graph $\Ga$. 
Then $G'$ takes the form sketched in Figure \nextfig{\figufft} 
($(a)$ and $(c)$ are 
the cases where $m_\one$ and $n_\one$ are even, $(b)$ is the case 
where they are odd. The dots indicate possible additional lines, 
i.e.\ the lines drawn are the minimal number that has to be there.
All $k_\one$  lines
crossing the dashed line are of scale $j$, the others are of higher scale.
$k_\one \ge 5$ because $G_\one$ has incidence number six or more.

\herefig{fst3fig15}

The lines of a possible spanning tree for $G'$ are drawn fat in 
Figure $\figufft$. It is clear by inspection that the 
statement of the Lemma holds. Moreover, one of the lines $\vth_\one$ in the 
spanning tree is at scale $j$, and so are two lines generating loops that 
overlap on $\vth_\one$.
The cases in Figure $\figufft$ are equivalent to those of Figure
$\figuele$; we redrew the graphs to bring out the scale structure. 
\endproof

\Lem{\mskel}{\Lesty  Assume \ATwo{2,0}--\AFou.
Let $G$ be a skeleton graph with $2m$ external 
legs and vertices 
$v$ with an incidence number $2m_v \ge 4$. 
Let $\cT$ be a tree rooted at a fork $\ph$ and such that $\cT \sim G$.
Assume that there are $\ep>0$, $n_v \ge 0$ $n_\ell \ge 0$, and 
that for all $s \in \natz{2}$ there are $\xi_{v,s} > 0$  and 
$\bbbQ_{l,s} > 0$ such that 
$$
\abs{\cU_v}_{s} \le
\cases{ \xi_{v,s} M^{j_v(2-m_v-s)} \la_{n_v}(j_v,\ep) & 
if $v$ is a vertex of scale $j_v<0$ \cr
\xi_{v,s} & if $v$ is a vertex of scale $j_v =0$}
\eqn $$
(where $\la_n$ is the function defined in \queq{\LAN})
and that the propagators associated to the lines of $G$ satisfy
$$
\abs{D^\si S_\ell (p)} \le \bbbQ_{\ell,s} \la_{n_\ell}(j_\ell,\ep)
M^{-j_\ell (1+s)} \; 
\True{|ip_\zer -e(\p)| \in \lbrack M^{j_\ell -2}, M^{j_\ell}
\rbrack}
.\eqn $$
For a fork $f \in \cT$ define
$$\eqalign{
n_f & = \sum\limits_{\ell \in L(G_f)} n_\ell + 
\sum\limits_{v\in V(G_f)} n_v + 
\abs{\{ v\in V(G_f): 2m_v=4, v \hbox{ not scale zero } \}} \cr
& + \abs{\{ f'\in \cT : f' \hbox{ fork}, f' > f, G_{f'} \hbox{ four--legged},
\tilde G_{f'} (f') \hbox{ is \NOL }\}}
.\cr}\EQN\nfdef $$
Let $j<0$ and
$$
\cS(j,\cT,G) = \sum\limits_{J \in \cJ(\cT,j)} Val (G^J)
.\eqn $$
Then for $s \in \natz{2}$
$$\eqalign{
\abs{\cS(j,\cT,G)}_s \le \cM (s,G) \K_\zer^{|L(G)|}
\left(\nu_\one |j| M^j \log M \right)^{\gOL (\tilde G (\ph))}\;
M^{j(2-m-s)} \la_{n_\ph}(j,\ep)
\cr}\EQN\mskbou $$
with 
$$
\K_\zer = 8 \nu_\zer \left( \nu_\one \log M\right)^2
\eqn $$
and
$$
\cM(s,G) = \max\limits_T \sum\limits_{\si\in Z(s,T,G)} 
\prod\limits_{\ell\in L(G)} \bbbQ_{\ell,|\si_l|} 
\prod\limits_{v\in V(G)} \xi_{v,\si_v}
,\eqn $$
where $Z(s,T,G)$ is as in Lemma $\ssBound$. 
}

\Proof Recall that the value of $G^J$ is given by 
$$\eqalign{
Val(G^J)(r_\one, \ldots, r_{2m-1}) = \sum\limits_{(A_v)_{v\in V(G)}}
\int & \prod\limits_{\ell\in L(G)\setminus L(T)} 
\frac{dp_\ell}{(2\pi)^{d+1}} 
\prod\limits_{\ell\in L(G)} \left(S_\ell(p_\ell)\right)_{\al_\ell\al_\ell'}\cr
& \prod\limits_{v\in V(G)} \cU_v\left(p_\one^{(v)}, \ldots, p_{2m_v-1}^{(v)}, 
A_v\right)
\cr}\eqn $$
where the $p_k^{(v)}$ are fixed by the usual rules in terms of the $p_\ell$ 
and the external momenta $r_\one, \ldots, r_{2m-1}$, and where $A_v \in 
\{ \uparrow,\downarrow\}^{2m_v}$, and the $\al_\ell$ and $\al_\ell'$ are
determined by the $A_v$ of the adjacent vertices. 

Let $\tilde G = \tilde G(\ph)$, $\tilde V = V(\tilde G)$ and $\tilde L = 
L(\tilde G)$. We may assume that the spanning tree of $G$ is chosen such that
the subgraph $\tilde T$ that it induces in $\tilde G$ is also a tree, 
and hence a spanning tree of $\tilde G$. If $\tilde v \in \tilde V$
is the image of $G_f$ belonging to a fork $f$ with $\pi(t)=\ph$ under the 
projection from $G$ to $\tilde G$, then  
the vertex function associated to $\tilde v$ is 
$$
\tilde \cU_{\tilde v} = \sum\limits_{j'> j} 
\sum\limits_{J \in \cJ (\cT,j')} Val (G_f^J)
.\EQN\pifbd $$
If $\tilde v$ is a vertex $\tilde v = v\in V(G)$ with $\pi (v) = \ph$
(i.e.\ there is no fork $f > \ph$ such that $v \in G_f$) and if 
$v$ carries scale $j_v < 0$, then 
$$
\tilde \cU_{\tilde v} = \sum\limits_{j_v> j} \cU_v
.\EQN\vsum $$
If $\tilde v$ is a scale zero vertex, then 
$\tilde \cU_{\tilde v} = \cU_v$.
With these definitions, 
$$
\cS (j,\cT,G) = Val (\tilde G)
.\EQN\GGt $$
To prove the Lemma, we do an induction in the height of $\cT$, defined as
$$
h(\cT) = \max\{k: \exists \hbox{ forks }f_\one, \ldots, f_k \in \cT, 
\hbox{ such that } f_k = \ph \hbox{ and } \pi(f_l)=f_{l+1} 
\hbox{ for all } l \in \nat{k-1}\}
.\eqn $$
If $h(\cT) =0$, $G=\tilde G(\ph)$, and $j_\ell = j$ for all $\ell \in L(G)$.
Then every vertex function $\tilde \cU_v$ is a scale sum \queq{\vsum} or 
given by $\cU_v $ for a scale zero vertex.  In the latter case, 
$$
\abs{\tilde \cU_{\tilde v} }_s \le \xi_{v,s} \le \xi_{v,s} M^{j(2-m_v-s)}
\eqn $$
since $2-m_v-s \le 0$ for all $s\ge 0$ and all $v$. In the former case,

\leftit{} if $m_v=2$ and $s=0$, 
$$\eqalign{
\abs{\tilde \cU_{\tilde v} }_\zer & \le \xi_{v,0} \sum\limits_{j_v> j}
\la_{n_v} (j_v,\ep) \le \xi_{v,0} (|j|+1) \la_{n_v} (j,\ep) \cr
& \le \xi_{v,0} \la_{n_v+1} (j_v,\ep)
\cr}\eqn $$

\leftit{} if $m_v\ge 3 $ or $s> 0$, we have $2-m_v-s < 0$, so 
$$\eqalign{
\abs{\tilde \cU_{\tilde v} }_s & \le \xi_{v,s} \sum\limits_{j_v> j}
\la_{n_v} (j_v,\ep) M^{j_v (2-m_v-s)} \cr
& \le \xi_{v,s} \la_{n_v} (j,\ep) M^{j(2-m_v-s)}
\sum\limits_{k=1}^\infty M^{-k(2-m_v-s)} \cr
&\le \frac{1}{M-1}\xi_{v,s} \la_{n_v} (j,\ep) M^{j(2-m_v-s)}
.\cr}\eqn $$
\pni
Thus the vertex functions of $\tilde G$ satisfy 
$$
\abs{\tilde \cU_{\tilde v} }_\si \le 
\xi_{v,s} \la_{\hat n_v} (j,\ep) M^{j(2-m_v-\si)}
\EQN\pivbd $$
with 
$$
\hat n_v = \cases{ n_v+1 & if $m_v=2$ and $v$ is not a scale zero vertex\cr
n_v & otherwise.\cr}
\eqn $$
By Lemma $\ssBound$, and since $j_\ell=j$ for all $\ell \in L(G)$, 
$$\eqalign{
\abs{Val(\tilde G)}_s \le\;  & 2 \; (4 \nu_\zer)^{|L(G)|} \; 
M^{j(2-m-s)} \left(\nu_\one |j| M^j \log M \right)^{\gOL(\tilde G)} 
\;  \cM(s,G)\cr
& \prod\limits_{\ell\in L(G)} \la_{n_\ell} (j,\ep) 
\prod\limits_{v\in V(G)} \la_{\hat n_v}(j,\ep) 
.\cr}\eqn $$
The product over the factors $\la_n$ is bounded by $\la_N(j,\ep)$, 
where, by \queq{\prinq},
$$
N=
|\{ v\in V(G): \pi(v)=\ph, m_v=2, v \hbox{ not scale zero }\}
 + \sum_{\ell\in L(G)} n_\ell + \sum_{v\in V(G)} n_v 
.\eqn $$
Since $h(\cT)=0$, there are no forks $f>\ph$ on $\cT$, so there are 
in particular no four--legged forks. Therefore $N=n_\ph$, and the statement
of the Lemma is proven for $h(\cT)=0$. 
\sni
Let $h(\cT) >0$ and \queq{\mskbou} be proven for all $\cT'$ with 
$h(\cT') \le h(\cT) -1$ and all skeletons $G'$ with $\cT' \sim G'$. 
Again, we first bound the vertex functions $\tilde \cU_{\tilde v}$  associated
to the vertices $\tilde v $ of $\tilde G$. If $\tilde v= v \in V(G)$
with $\pi(v)=\ph$, the bound \queq{\pivbd} holds. If $\tilde v$ belongs 
to a fork $f$ with $\pi(f)=\ph$, the subtree $\cT_f$ of $\cT$ rooted at $f$ and 
$G_f$ fulfil the induction hypothesis. Using \queq{\pifbd} and bounding the 
scale sums as in the proof of \queq{\pivbd}, we get
$$\eqalign{
\abs{\tilde \cU_{\tilde v}}_\si \le  & \cM (\si,G_f) \; 
(\nu_\one \log M ) \; \K_\zer^{|L(G_f)|} 
\la_{\hat n_f}(j,\ep) M^{j(2-m_f-\si)}
\cr}\EQN\tUtv $$
where $2m_f$ is the number of external legs of $G_f$ and
$$
\hat n_f = \cases{ n_f+1 & if $2m_f=4$ and $\tilde G_f(f)$ is \NOL\cr
n_f & otherwise.}
\eqn $$

All lines of $\tilde G$ have scale $j_\ell =j$. Lemma $\ssBound$ applies 
to $\tilde G$. Thus, by \queq{\GGt}, calling $R(M)=\nu_\one\log M$, 
$$
\abs{\cS(j,\cT,G)}_s \le \; 2 (4\nu_\zer)^{|\tilde L|} \; 
M^{j(2-m-s)} (|j|M^j \nu_\one \log M)^{\gOL(\tilde G)}
\; \La \; \Ga \; \Si
\eqn $$
with 
$$\eqalign{
\La & = \prod\limits_{v\in V(G) \atop \pi(v)=\ph} \la_{\hat n_v}(j,\ep)\;
\prod\limits_{f\in \cT\atop\pi(f)=\ph} \la_{\hat n_f}(j,\ep)\;
\prod\limits_{\ell \in \tilde L} \la_{n_\ell}(j,\ep)
\le \la_{n_\ph}(j,\ep)\cr
\Ga &=  \prod\limits_{v\in V(G) \atop \pi(v)=\ph}  R(M) \; 
\prod\limits_{f\in \cT\atop\pi(f)=\ph} R(M) \K_\zer^{|L(G_f)|} \cr
\Si &= \sum\limits_{\si\in Z(s,\tilde T,\tilde G} \; 
\prod\limits_{\ell \in \tilde L} \bbbQ_{\ell, |\si_\ell|}\;
\prod\limits_{v\in V(G) \atop \pi(v)=\ph} \xi_{v,|\si_v|}\;
\prod\limits_{\tilde v = f\in \cT\atop\pi(f)=\ph}
\cM(|\si_{\tilde v}|,G_f)
\cr}\eqn $$
To combine the constants, we use that 
$$
L(G) = L(\tilde G) \dotcup \bigcup\limits_{f:\pi(f)=\ph}^\cdot L(G_f)
\eqn $$
and 
$$
V(G) = \{ v\in V(G): \pi(v)=\ph\} \dotcup 
\bigcup_{f: \pi(f)=\ph}^\cdot V(G_f)
.\eqn $$
Thus 
$$\eqalign{
\prod\limits_{v\in V(G) \atop \pi(v)=\ph} R(M)
\prod\limits_{f\in \cT\atop\pi(f)=\ph} R(M) & \le
R(M)^{|\tilde V|} \le R(M)^{|\tilde L|+1} 
\le R(M)^{2|\tilde L|}\cr
\prod\limits_{f\in \cT\atop\pi(f)=\ph} \K_\zer^{|L(G_f)|} 
& = 
\K_\zer^{|L(G)|-|\tilde L|}
,\cr}\eqn $$
so  
$$
2 (4\nu_\zer)^{|\tilde L|} \;\prod\limits_{f\in \cT\atop\pi(f)=\ph}
\K_\zer^{|L(G_f)|} \prod\limits_{v\in V(G) \atop \pi(v)=\ph} R(M)
\prod\limits_{f\in \cT\atop\pi(f)=\ph} R(M) \le \K_\zer^{|L(G)|}
.\eqn $$
Finally, we bound $\Si$. Every $\cM(|\si_f|,G_f)$ in the product
is a maximum over spanning trees of $G_f$, which is attained at 
some $T_f$ since the set of spanning trees is finite. The union of all such 
$T_f$ and $\tilde T$ is a spanning tree $T^*$ of $G$. The sum over the 
multiindices of the $G_f$ and those of $G$ combines to a sum over
$\si\in Z(s,T^*,G)$. Collecting the $\bbbQ_{\ell,|\si_\ell|} $ and 
$\xi_{v,|\si_v|}$,  we obtain
$$
\Si = \sum\limits_{\si\in Z(s,T^*,G)} 
\prod\limits_{\ell \in L(G)} \bbbQ_{\ell, |\si_\ell|}\;
\prod\limits_{v\in V(G) } \xi_{v,|\si_v|}\;
.\eqn $$
This sum is bounded by its maximum over all possible spanning trees of 
$G$, which is $\cM(s,G)$. \endproof

\def\tG{{\tilde G}}
\The{\twoDOL}{\Lesty Assume \ATwo{2,\hxp}--\AFou\ and 
let $G$ be a two--legged skeleton graph obeying
the hypotheses of Lemma $\mskel$. Let $t_{\tG(\ph)} \ge 1$, and 
$\tG(\ph)$ not be a sunset (Figure $\dolone$) or a \WL\ (Figure $\wicked$). 
Let all vertices $v$ have scale $j_v=0$, and
$$
|\cU_v|_s \le \xi_{v,s} \quad \left( = \xi_{v,s} M^{j_v(2-m_v-s)}\right)
.\EQN\whatelse $$
Let 
$$
\cS (j,\cT,G) = \sum_{J \in \cJ(\cT,j)} Val(G^J)
.\EQN\thatscS $$
Then for $s\in \{ 0,1,2\}$, 
$$
\abs{\cS (j,\cT,G)}_s \le \K_\zer^{|L(G)|} |j|^3 M^{j(3-s)}
\cM(s,G)\la_{n_\ph}(j,\ep)
,\EQN\DOLc $$
with $\cM$ and $\K_\zer$ given as in Lemma $\mskel$.
}

\Proof  Let $\tilde G = \tilde G(\ph)$. Then $\tilde G$ is a skeleton 
or it consists of one external vertex only, with self--contractions. 
$\tilde G$ has vertices $\tilde v$ that carry scales $j_{\tilde v}$,
where $j_{\tilde v}=0$ for scale zero vertices, or $j_{\tilde v}>j$
for a vertex belonging to a fork $f\in \cT$ with $\pi(f)=\ph$. 
In the latter case, the vertex function is 
$$
\cU_{\tilde v} = \sum_{J\in \cJ (\cT_f,j_{\tilde v})} Val(G_f^J)
.\EQN\summedvert $$
If $\tG$ is not a \MS, then $\tG$ is DOL by Theorem \allnonDOL. 
By Lemma \mskel, 
$$
\abs{\cS(j,\cT,G)}_s \le \cM(s,G) \K_\zer^{|L(G)|} 
(\nu_\one \log M)^2 |j|^2 M^{2j} M^{j(1-s)} \la_{n_\ph}(j,\ep)
\eqn $$
which proves \queq{\DOLc}. 
So let $\tG$ be a \MS. There are three cases.
\ssni
1. One of the vertices, say $v_\one$, is of scale zero. 
Let the number of lines between $v_\one $ and $v_\two$ be $n$, 
then $n\ge 5$, and $G$ has  $n-1$ loops. All $j_\ell=j$. 
$\gOL(\tG)=1$. 
Since all vertices of the graph $G$ are scale zero, the scale-dependent 
vertex $v_\two $ of $\tG$ corresponds to a fork $f$ with a subgraph $G_f$.
Lemma \mskel\ applies to $G_f$, so the vertices of $\tG$ satisfy 
$$\eqalign{
\abs{\cU_{v_\one}}_s &\le \xi_{v,s} \qquad s \in \{ 0,1,2\}\cr
\sum_{j_{\tilde v} > j} \abs{\cU_{v_\two}}_s &\le R(M) \cM(s, G_f) \K_\zer^{|L(G_f)|} 
\la_{n_f} (j,\ep) M^{j(2-m_f-s)} \qquad s \in \{ 0,1,2\}
.\cr}\eqn $$
Let $\ell_\one \in L(\tG)$ be 
the line in the spanning tree, then 
$$\eqalign{
\abs{\cS(j,\cT,G)}_s = \abs{Val(\tG)}_s & \le 
\sum_{\si_{v_\one}+\si_{v_\two}+\si_{\ell_\one} \le s}
\abs{\cU_{v_\one}}_{|\si_\one|} \; 
\left(\sum_{j_{\tilde v} > j} \abs{\cU_{v_\two}}_{|\si_\two|} \right) \; 
\bbbQ_{\ell_\one,|\si_{\ell_\one}|} \ M^{-j(1+|\si_{\ell_\one}|)}\cr
& \left(\prod\limits_{\ell\neq\ell_\one} \bbbQ_{\ell,0}\right) M^{-(n-1)j}
(2 \nu_\zer M^{2j})^{n-1} \; \nu_\one M^j |j| \log M \cr
 &\le \sum_{\si_\one+\si_\two+\si_\thr \le s}
\xi_{v_\one,|\si_\one|}\cM(|\si_\two|,G_f) \K_\zer^{|L(G_f)|} 
\la_{n_f} (j,\ep) M^{j(2-m_f-|\si_\two|)} \cr
& (2\nu_\zer)^{n-1} (\nu_\one |j|\log M) M^{j(n-2+1-|\si_\thr|)} 
\bbbQ_{\ell_\one,|\si_\thr|} \prod\limits_{\ell\neq\ell_\one} \bbbQ_{\ell,0}
.\cr}\eqn $$
We use $m_f=\sfrac{n+1}{2}$ and $|\si_\two|+|\si_\thr| \le s$ to get
$$\eqalign{
\abs{\cS(j,\cT,G)}_s  &\le M^{j(\sfrac{n+1}{2}-s)} |j| 
\sum_{\si_\one+\si_\two+\si_\thr \le s}
\xi_{v_\one,|\si_\one|}\cM(|\si_\two|,G_f) \K_\zer^{|L(G_f)|} \cr
& \qquad \la_{n_f} (j,\ep) (2\nu_\zer)^{n-1} (\nu_\one \log M) 
\bbbQ_{\ell_\one,|\si_\thr|} \prod\limits_{\ell\neq\ell_\one} \bbbQ_{\ell,0}\cr
& \le  M^{j(\sfrac{n+1}{2}-s)} |j| \K_\zer^{|L(G)|}\la_{n_\ph} (j,\ep)
\cM(s,G)
.\cr}\eqn $$ 
Since $n\ge 5$, \queq{\DOLc} follows.
If $v_\one $ comes from a summed vertex, the same bound holds.
\ssni
2. Both vertices of $\tG$ are of scale zero. The bound 
in 1.\ improves 
by a factor $M^{j(\sfrac{n+1}{2})-2}$ (see the discussion 
around \queq{\AroundEq}. 
\ssni
3. Both vertices $\tilde v_\one $ and $\tilde v_\two$ are scale-dependent.
Since $G$ has only scale zero vertices, the $\tilde v_i$ must belong to 
subgraphs of $G$. If all lines of these subgraphs are self--contractions
of scale zero vertices, the bound is similar to that of case 2. 
So we may assume that $G$ has at least three vertices.
Therefore Lemma \Fnine\ applies. Let $\cT'$ be as constructed in Lemma 
\Fnine, and let $G'=\tG(\cT')$, so that 
$$
Val(\tG) = \sum_{J\in \cJ (\cT',j)} Val ({G'^J})
.\eqn $$
The vertex functions for all $v' \in G'$ are given by scale sums 
$\cS (j_f,\cT_f,G_f) $ of skeleton graphs $G_f$ with $\pi(f)=f'$,
where $f'$ is the highest fork of $\cT$ that is in $\cT'$. 
$j_f$ is summed down to $j'=j_{f'}$. The $\sum\limits_{j_f > j'}
\cS (j_f,\cT_f,G_f) $ therefore obey
\queq{\mskbou}, with $\gOL=0$ and $j$ replaced by $j'$. 
We choose a suitable spanning tree
$T^*$ of $G'$, as in Lemma \Fnine. Derivatives of $Val(G'^J)$ are given 
by \queq{\DbeValG}, and can be bounded as in \queq{\fthn} and \queq{\ffty},
with $\cV(G,T)$ replaced in \queq{\fthn} by the integration volume 
$$\eqalign{
\cV'(G',T^*) = \int &\prod\limits_{\ell \in L(G')\setminus L(T^*)} 
\left( d^{d+1} p_\ell \; \True{|i(p_\ell)_\zer - e(\p_\ell)| \le M^{j_\ell}}\right)
\cr
& \prod\limits_{\be \in L(T^*)} 
\True{|i(p_\be)_\zer - e(\p_\be)| \le M^{j_\be}}
.\cr}\eqn $$
Doing the $p_\zer$--integrals, we get
$$
\cV'(G',T) \le \bar\cV(G',T^*) \prod\limits_{\ell\in  L(G)\setminus L(T^*)}
(2M^{j_\ell})
,\eqn $$
with $\bar \cV$ as in Lemma \DOLVolnoamol.
$G'$ is DOL, and by Lemma \Fnine, one of the volume gains arises at 
scale $j$, and the other one at scale $j'$, which is the lowest scale 
on which a derivative can act in $G_{f_\one}$. By Lemma \DOLVolnoamol, 
and since $j_\ell \ge j$ for all $\ell \in L(G')$, 
$$
\cV'(G',T^*) \le 
(\nu_\one |j|)^2 \; M^{j+j'} 
\prod\limits_{\ell\in  L(G)\setminus L(T^*)}
(2\nu_\zer M^{2j_\ell})
.\eqn $$
Also, 
$$\eqalign{
Y(\be,T^*) & = \sum_\si 
\left(\prod\limits_{\ell \in L(G')} \bbbQ_{\ell, |\si_{\ell,r}|}
\la_{n_\ell}(j_\ell,\ep) 
M^{-j_\ell (1+|\si_{\ell,r}|)}\right)
\prod\limits_{v\in V(G')} \abs{\cU_v}_{|\si_{v,r}|} \cr
& \le \sum_\si
\left(\prod\limits_{\ell \in L(G')} \bbbQ_{\ell, |\si_{\ell,r}|}
\la_{n_\ell}(j_\ell,\ep) M^{-j_\ell (1+|\si_{\ell,r}|)} \right) \cr
& \prod\limits_{v\in V(G')} \cM (|\si_{v,r}|,G_v) 
\K_\zer^{|L(G_v)|} M^{j_v(2-m_v-|\si_{v,r}|)} \la_{n_v}(j_v,\ep)
.\cr}\eqn $$
We bound $M^{-j_\ell |\si_{\ell,r}|} \le M^{-j |\si_{\ell,r}|}$ 
and $M^{-j_v |\si_{v,r}|} \le M^{-j |\si_{v,r}|}$, and use
$$
\sum |\si_{\ell,r}| + \sum |\si_{v,r}| \le s
.\eqn $$
The scale factors without the improvement factor $M^{j+j'}$
and the derivative factor $M^{-sj}$ add up to
$$
B=-\sum_{\ell \in L(G')} j_\ell + \sum_{v\in V(G')} 
j_v (2-m_v) + \sum_{\ell \in L(G')\setminus L(T^*)} 2 j_\ell
.\eqn $$
Since 
$$
j_\ell = j+ \sum\limits_{\ph < f \in \cT'\atop \ell \in G_{f'}}
(j_f - j_{\pi(f)})
\eqn $$
and (since the scale sum over the effective vertices is already done)
$$
j_v = j+ \sum\limits_{\ph < f \in \cT'\atop v \in G_{f'}}
(j_f - j_{\pi(f)})
\eqn $$
so 
$$\eqalign{
B &=j\left(\sum_{v\in V(G')} (2-m_v) + |L(G')| - 2 |L(T^*)|\right) \cr
&+\sum\limits_{f>\ph} (j_f - j_{\pi(f)}) 
\left( \sum_{v\in V(G_f')} (2-m_v) + |L(G_f')| - 
2 |L(G'_f)\cap L(T^*)|\right) 
.\cr}\eqn $$
Using $|L(G'_f)| = \sum\limits_{v\in V(G_f')} m_v \; -m$ and 
$ |L(G'_f)\cap L(T^*)|= |V(G'_f)|-1$, we get
$$
B= j(2-m) + \sum_{\ph < f \in \cT'} (j_f - j_{\pi(f)}) (2-m_f)
.\eqn $$
By the structure of $\cT'$ given in Lemma \Fnine, 
$$
B= j(2-m) + \sum_{f_\one \le f \le f'} (j_f - j_{\pi(f)}) (2-m_f)
,\eqn $$
so, adding $j$ and 
$j' = j+ \sum\limits_{f_\one \le f \le f'} (j_f - j_{\pi(f)})$,
we get 
$$
j+j'+B=j(4-m) +  \sum\limits_{f_\one \le f \in \cT'} (j_f - j_{\pi(f)})
(3-m_f)
.\eqn $$
Using $m=1$ and adding the scale effect $-sj$ from $s$ derivatives, 
we have 
$$
j+j'-sj+B=j(3-s) +  \sum\limits_{f_\one \le f \in \cT'} (j_f - j_{\pi(f)})
(3-m_f)
.\eqn $$
By Lemma \Fnine, only $f_\one$ can have $m_f =3$, all higher forks of 
$\cT'$, if existent, must have $3-m_f \le -1$. 
Thus, combining the constants, 
$$\eqalign{
\abs{Val(\tG)}_s &\le M^{j(3-s)} \la_{n_\ph}(j,\ep)
|j|^2 \cM(s,G') \K_\zer^{|L(G')|}\cr
&\qquad \sum\limits_{(j_f)_{f\in \cT'\setminus \{ \ph\}}}
\prod\limits_{f\in \cT'\setminus \{ \ph\}}
M^{(3-m_f) (j_f-j_{\pi(f)})} \cr
& \le |j|^3 M^{j(3-s)} \la_{n_\ph}(j,\ep) \cM(s,G) \K_\zer^{|L(G)|}
.\cr}\eqn $$
\endproof 

\Rem{\aboutHyp} None of the results of this section required \AFiv\ 
because only volume gains were used in the bounds. \AFiv\ is
only necessary to control the contributions from RPA graphs.

\sect{RPA skeletons and general graphs in two dimensions}

\Def{\genRPA} Let $G$ be a two--legged graph and $\cT$ be a tree
compatible to $G$, rooted at $\ph$. 
The graph $G$ is a generalized RPA graph (for tree $\cT$)
if $\tG(\ph)$ is a sunset or a wicked ladder.
\ssni
To bound the values of generalized RPA graphs 
(not covered in Theorem \twoDOL), we shall need \AFiv. The proof
of regularity is by reduction to DOL graphs at higher scales, 
or by reduction to the cases treated in \TWO. In the proof of 
the following theorem, we shall go through a number of easy arguments
to do this reduction. 

\The{\twononDOL}{\Lesty  Assume \ATwo{2,\hxp}, \AThr, and \AFou. 
Let $G$ be a two--legged graph without  
two--legged proper subgraphs and $\cT$ be a tree
compatible to $G$, rooted at $\ph$,
such that $\tG=\tG(\ph)$ is a sunset, or a \WL,
or has only one external vertex. Let $G$ have vertices 
$v$ with an incidence number $2m_v \ge 4$, all of scale zero,  
with $C^2$ vertex functions $\cU_v$ obeying \queq{\whatelse},
and with propagators satisfying \queq{\Sgutt} associated to the lines.
Then $\cS (j,\cT,G)$, defined in \queq{\thatscS}, satisfies
for $s=1$ and $s=2$
$$
\abs{\cS (j,\cT,G)}_s \le \K_5(G) |j| M^{(2-s)j}
(\nu_\one \log M) \la_{n_\ph} (j,\ep)
.\EQN\OLc $$
Assume \AFiv\ and let the propagators associated to lines be
given by the strings $S_{j_\ell}$ of \queq{\soeinstring} 
satisfying \queq{\Sgutt } and \queq{\thsok}.
Then the projection $\ell$ onto the Fermi surface satisfies 
$$
\abs{\ell \cS(j,\cT,G)}_\two \le \K_6(G) |j| M^{j/3} \la_{n_\ph} (j,\ep)
,\EQN\hart $$
and
$$
\abs{\frac{\del}{\del\th} \frac{\del}{\del p_\al}
 \cS(j,\cT,G)}_\zer \le \K_7(G) |j| M^{j/3} \la_{n_\ph} (j,\ep)
\EQN\haerter $$
}

\Proof Let $\tilde G = \tilde G(\ph)$. Consider the case in 
which $t_\tG\ge 1$. The case when $t_\tG = 0$ will be considered
at the end of the proof.
Now $\tilde G$ is a skeleton. 
$\tilde G$ has vertices $\tilde v$ that carry scales $j_{\tilde v}$,
where $j_{\tilde v}=0$ for scale zero vertices, or $j_{\tilde v}>j$
for a vertex belonging to a fork $f\in \cT$ with $\pi(f)=\ph$. 
In the latter case, the vertex function is as in \queq{\summedvert}.
If $\tG$ is a sunset or a \WL, then $\gOL(\tG)\ge 1$, so by Lemma \mskel\
$$
\abs{\cS(j,\cT,G)}_s \le \cM(s,G) \K_\zer^{|L(G)|} 
(\nu_\one \log M) |j| M^{j} M^{j(1-s)} \la_{n_\ph}(j,\ep)
\eqn $$
which proves \queq{\OLc} and the bounds on the $0^{th}$ and the
first derivatives required for \queq{\hart}. 
It remains to bound the second derivative contributions to \queq{\hart}
and \queq{\haerter}.
\ssni
1. \hroom 
Let $\tG$ be a sunset.  Then there are three possibilities for $G$ itself:
$G$ can be a sunset, in which case $G=\tG$, or $G$ is a \WL, or $G$ is DOL.
\ssni
1.1 \hroom
$G$ a sunset. The vertices of $G$ may have more than four legs and 
have self--contractions, but that is inessential for the following 
discussion since the external momentum cannot enter those `tadpole lines', 
and because the scale behaviour of a loop containing exactly one 
fermion line is $M^{-j} \; M^{2j} = M^j$, which is summable and produces
the same bound as if the self--contraction were not there.
We may therefore assume that there are no self--contractions. The projection 
$\ell$ means evaluation on the Fermi surface, $(\ell F) (p) = 
F(0,\p(0,\th))$, so the only derivatives we look at are those with 
respect to $\th$. 

The proof is an application of Theorem \dreipro. 
Unlike $C_j(p_\zer,e(\p))$, the propagators
$S_{\ell,j_\ell}(p)$ depend on $\th$ because the $T_k^{(j)}$
from \queq{\soeinstring} do. However, \queq{\thsok} ensures that the same
change of variables used in \TWO\ will prevent the derivative from 
degrading the scale behaviour here too. We use the notation
of \TWO. The function $P$ (see \queq{\voilaP}) entering \queq{\YUOm}
still contains scale zero vertex functions only, hence is $C^2$ with 
bounds uniform in the scales. 
We now have instead of \queq{\Yuljpi}
$$
Y_{j,a}(p) = \int dp_\one \int dp_\two \; 
S_{\ell_\one,j}(p_\one)\;
S_{\ell_\two,j}(p_\two) \;
S_{\ell_\thr,j}(L_a(p_\one,p_\two,p))\; 
P(p_\one,p_\two,p)
\eqn $$
with $\ze_a$ and $e_a$ given by \queq{\eazea}. All $S_{\ell_k,j}$ are 
$C^{2,\hxp}$.  We take one derivative with respect 
to $p$ right away. In the integrand, it acts only on 
$S_{\ell_\thr,j}$ and on $P$. Thus $\sfrac{\del}{\del p}Y_{j,a}(p)$
consists of two terms, both of the form given in \queq{\Xuljbulnudef}.
In the first term, where the derivative acts on $P$, $A=\sfrac{\del}{\del
p}P$, and $\nu_\one=\nu_\two=\nu_\thr=1$. In the second term,
where the derivative acts on $S_{\ell_\thr,j}$, $\nu_\one=\nu_\two=1$,
and $\nu_\thr=2$. Both terms fulfil the hypotheses of Theorem
\dreipro. Thus, by Theorem \dreipro, 
$$
\abs{\frac{\del}{\del \th}\; \frac{\del Y_{j,a}}{\del p}}_\zer
\le \Const \abs{j_\one} M^{j_\one+j_\two-{3\over 2} j_\thr}
\eqn $$
with the constant given as in Theorem \dreipro. This implies
convergence of the scale sums, as shown in all detail in 
\TWO, Section 3.2.

\ssni
1.2 \hroom $G$ a \WL. 
\ssni
1.2.1 \hroom \SYmm\ does not hold, and $G$ is a particle--particle \WL.
Then we can use an argument as in the proof of Theorem \goodWL\ 
to show that the extra volume gain from Lemma \bubvol\ suffices to 
make the second derivative convergent. Note that although not all 
lines have the same scale, the spanning tree can always be chosen
consistent with the scales such that both the gain of Lemma \bubvol\
and the gain from a pair of \OL\ loops can be extracted (the latter on 
root scale $j$). An example is the graph shown in Figure 
\nextfig{\noabuidl}. The line connecting the two external 
vertices carries scale $j$ because the $\tG(\ph)$ is a sunset.
If, for instance, $j_\two \ge j_\thr$ and $j_5 \ge j_4$,
one puts the lines with scales $j_\two$ and $j_5$ into the 
spanning tree, as indicated by the heavy lines in the Figure.
The details of the volume bound are given in Lemma \TWO.4.8 and 
Theorem \TWO.4.9. In this example, there are two volume gains
from particle--particle bubbles and one gain from overlapping
loops on scale $j$. The total improvement is $M^{j+{j_5+j_2\over
3}}$ by Lemma \TWO.4.8. This improvement is used to control 
derivatives that act strictly above root scale. Those that
act at root scale are controlled using the change of variable
arguments of Theorem \TWO.3.5.

\herefig{fst3fig16}

\ssni
1.2.2 \hroom If \SYmm\ holds or $G$ is not a particle--particle
\WL, 
we choose the spanning tree for $G$ as in Figure $\TWO.6$.
For the same reasons as in case 1.1 (namely \queq{\thsok}), 
we may apply the same change of 
variables procedure as for the \WL\ with propagators $C_j$. 
It is no complication that the lines in different bubbles may 
be of different scale since the line with $p_\one$ is always 
on scale $j$. However, even the two lines in a single bubble
may now have different scales, and in particular, the line in the tree 
may be the one with the lower scale. In the particle--particle \WL, 
we may simply change the spanning tree to contain the higher of 
the two lines since this amounts to a change of variables from 
$\p_\thr$ to $\q_\thr = \p+\p_\one - \p_\thr$, thus
$\p_\thr=\p+\p_\one-\q_\thr$, which does not change any signs 
and hence also no critical points.
In the particle--hole \WL,  the change of the spanning tree only 
exchanges the signs in front of $\p_\one$ and $\p$, and the critical 
points do not change because of that. Thus the procedure of Section 
\TWO.3.5 applies directly.
\ssni
1.3 \hroom $G$ DOL. We construct a maximal non--DOL quotient graph $\tG(\cT')$
of $G$, associated to a tree $\cT'$, which is grown as follows. 
For both effective vertices of $\tG=\tG(\phi)$, 
associated to forks $f_\one$ and $f_\two$, we first determine if 
$\tG_i=\tG_{f_i}(f_i)$ are \OL. If $\tG_i$ is \NOL, we add $f_i$ to the 
tree $\cT'$. We now attempt to add each fork $f$ associated to
effective four-legged vertices of the current quotient graph.
After each trial addition we check to see
if the resulting $\tG(\cT')$ is DOL. This happens, for example, if the graph
in Figure \figutwe\ $(b)$ is expanded to the one in Figure 
$\figuthi$. In this case, the fork $f$ is rejected.
Otherwise, $\tG(\cT')$ must be a \WL, and the fork is appended to $\cT'$. 
Once all candidate forks are rejected, $\cT'$ is complete. 
As a result, 
$\Ga$ is a \WL, with vertices that are of scale zero or 
belong to \OL\ subgraphs, or $\Ga$ becomes a DOL graph $\tilde
\Ga$ when one of its
nonoverlapping four-legged effective vertices is expanded.
In the former case, 1.2.1 and 1.2.2 apply, 
the only change being that there are now
vertex functions that depend on the scales. However, they come from 
\OL\ graphs, so whenever derivatives act on them, there is also one 
volume gain that improves the scaling behaviour by $M^{j} |j|$. 
In the latter case, we apply the procedure 
for the \WL s to $\Ga$. Whenever a derivative hits a vertex
of $\Ga$ , the corresponding subgraph is \OL, or the derivative 
acts on a line of $\tilde \Ga$ at the scale where the second overlap takes place.
Therefore, its effect is controlled by the volume gain at that scale. 
\ssni
2. \hroom Let $\tG$ be a \WL. The proof is the same as in case
1.2 and 1.3. One only has to construct $\cT'$ before applying
the results of Chapter \TWO.4, to ensure that the effective vertex
functions appearing in the \WL\ $\tG(\cT')$ are either scale
zero vertices or values of \OL\ graphs.
\sni
The remaining case is $t_\tG=0$. There is the trivial case, where 
a scale zero vertex has only self--contractions. Then $\cS(j,\cT,G)$
obeys the desired bounds by $(H1)$ and the properties of the scale zero action.
Otherwise,  $G$ is \OL. 
We consider the minimal quotient graph $\Ga $ of 
$G$ that has $t_\Ga\ge 1$. All the considerations of the above case apply,
the only difference being that the lowest scale on which the derivatives
can act is a scale strictly above root scale, not root scale itself, 
which just improves the estimates. 

\endproof

\Rem{\dolly} Let $\ga \in (0,1)$ and $G$ be a 1PI graph.
If \AOne{2,\ga}, \ATwo{2,0}, \AThr, and either \SYmm\
or \AFou\ hold, and if the pair $(\cT,G)$ is such that 
$\tG(\ph)$ is DOL, then the scale sum 
$$
\Si_{\rm DOL}(\cT,G,p) =\sli_{j<0} \sli_{J \in \cJ(\cT,j)} Val(G^J) (p)
\eqn $$
is $C^{2,\ga}$ in $p$. In other words, {\it $\Si_{\rm DOL}(\cT,G)$ 
is more regular than $e$}. 
The reason for this is that the double volume gain on root scale suffices 
to control almost three derivatives,
and that for 1PI graphs, one can always use an integration
by parts to remove a derivative from a propagator, so that one
can distribute the derivatives such that at most one derivative
and one difference operator can act on any given line (see Case 6 
of the proof of Theorem \ONE.2.46). Thus at most the second derivative
of $e$ appears in the bound.

\The{\ctwoness}{\Thsty Let $d=2$, 
assume \AOne{2,\hxp}, \ATwo{2,\hxp}, and \AThr--\AFiv, 
and let $K^I_r$ be the $r^{\rm th}$ order coefficient of $K^I$
i.e., $K^I = \sli_{r=1}^\infty K^I_r \la^r$ as a formal power
series in $\la$. Then
$$
\abs{K^I_r}_\two \le Q \; |\la|^r
.\eqn $$
The constant $Q$ depends only on $r$, $g_\zer$, $r_\zer$, and
$w_\zer$. As $I\to -\infty$, $K_r^I$ converges in 
$\abs{\cdot}_\two$ to a $C^2$ function $K_r$.  }
\Rem{\dummy} The $r$--behaviour is 
$$
Q = \tilde Q^r \; r!
\eqn $$
where $\tilde Q$ depends on $g_\zer$, $r_\zer$, and $w_\zer$. 

\Proof Let $G$ be a graph contributing to $K^I_r$. 
Then $G$ is two--legged and 1PI (otherwise $\ell$
of its value would be zero).
The contribution of $G$ to $K^I_r$ can be written as a scale sum 
$\sum_{j=I}^{-1} \sum_\cT \cS(j,\cT,G)$, 
where $\cS$ denotes the value of the graph
(see \ONE.2.76\ for the detailed formula for $K_r^I$). 

The statement of the Theorem holds for $r=1$ and $r=2$
by Theorem \zweireg. We do an induction in $r$. 
The inductive hypothesis consists of
\queq{\OLc}, \queq{\hart}, and \queq{\haerter}.
Let $r\ge 3$, assume the statement to be
proven for all $r' < r$, and let $G$ be a graph contributing
to $K^I_r$.  Let $G_\one$ be the skeleton graph
obtained from $G$ by replacing strings of two--legged subdiagrams
by single lines $l$, and associate the value $S_{l,j_l}$ of the string
to the line of $G_\one$. Every subdiagram is of order at most
$r-1$, so the inductive hypothesis applies to it. 
We now show that the values of the $c$-- and $r$--forks 
fulfil the hypotheses of Lemma \stringprop. 

\def\tT{\tilde T}
We start with the $c$--forks. They are given by a scale sum
$$
T^{(j)} (p) = \sli_{I \le h \le j} \ell \tT_h (p)
\eqn $$
with $\tT (p)$ the value of a two--legged 1PI graph $G_f$
corresponding to a fork $f$ of $\cT$. 
Since $\tT_h$ obeys \queq{\OLc}, and since the highest scale
in the sum appearing is $h=j$, \queq{\bravetes} is obvious 
for $s=0$ and $s=1$
(for details, see (\ONE.2.151-153)). Also, derivatives 
acting on $T^{(j)} (p)$ can only act as $\th$--derivatives. 
Similarly, \queq{\hart} and a summation over $h \le j$ imply
that $\abs{T^{(j)}}_\two \le \Const \la_{n_f} (j,\veps) M^{j/2}$,
which fulfils (and is actually much better than) 
\queq{\bravetes} with $s=2$.

If $T^{(j)} $ belongs to an $r$--fork, then 
$$
T^{(j)} (p_\zer,\p(\rh,\th)) = \sli_{h > j}
(1-\ell) \tT_h (p_\zer,\p(\rh,\th)) =
\sli_{h > j} \left(\tT_h(p_\zer,\p(\rh,\th))  - 
\tT_h (0,\p(0,\th))\right)
\eqn $$
so \queq{\bravetes} follows by the same scale summations and
estimates as in (\ONE.2.154--157). Moreover,
$$\eqalign{
\frac{\del}{\del\th} (1-\ell) T^{(j)} (p_\zer,\p(\rh,\th)) & = 
\sli_{h > j} \frac{\del}{\del\th} \left(\tT_h(p_\zer,\p(\rh,\th))  - 
\tT_h (0,\p(0,\th))\right)
\cr
&=\sli_{h > j} \left((\nabla \tT_h) (p_\zer,\p(\rh,\th)) \cdot
\frac{\del \p}{\del \th} (\rh,\th) - 
(\nabla \tT_h) (0,\p(0,\th)) \cdot
\frac{\del \p}{\del \th} (0,\th)\right)
\cr
&=
\left((1-\ell) (\nabla T^{(j)} \cdot \frac{\del \p}{\del \th})
\right) (p_\zer,\p(\rh,\th)) 
\cr}\eqn $$
In other words, the $\th$ derivative does not upset 
the renormalization cancellation. 
Taylor expanding, and inserting \queq{\OLc}, we
get 
$$\eqalign{
\abs{\frac{\del}{\del\th} (1-\ell) T^{(j)}}_\zer^{(j)} & \le
M^j
\sli_{h > j} \K_5(G_f) \; |h| \nu_\one \log M \; \la_{n_\ph} (h,\ep)
\cr
& \le \K_5(G_f) \nu_\one \log M \; \la_{n_\ph} (j,\ep) |j|^2 M^j
.\cr}\eqn $$
The second inequality in \queq{\braveretes} is proven by summing
\queq{\OLc}.

Thus $S_{l,j_l}$ fulfils the hypotheses of Lemma \stringprop,
and thus of Lemma \mskel, 
Theorem \twoDOL, and Theorem \twononDOL. 
By Theorem \twononDOL, the value of every graph
with root scale $j$ and GN tree $\cT$ contributing to $K_{r,j}^I$
($K^I_r=\sum\limits_{I \le j < 0} K_{r,j}^I$) satisfies
\queq{\OLc}, \queq{\hart}, and \queq{\haerter}.
Thus the scale sum over $j$ converges. The sum over trees gives
a constant to the power $r$. The proof of the $r!$--bound
is as in \quref{FT1}. 
\endproof

\Rem{\dummy} 
H\" older continuity of the second derivative follows by the
standard argument given in the Appendix and in Section \TWO.3.4 
(the hypotheses on the general propagators in Theorem \TWO.3.8 
are proven for strings in Lemma A.3).

\sect{Higher dimensions}
The detailed analysis of the preceding section, as well as
\AFiv, were necessary only
for $d=2$. For $d \ge 3$, the simpler argument of Section \TWO.3.5
applies to the generalized RPA graphs.
This is one point where one sees that the effective
strength of the infrared singularity is the same for all $d\ge 2$
only on the level of naive power counting. For improved power 
counting, which is necessary for taking derivatives, increasing the 
dimension helps. The geometrical reason for this is that although 
the codimension of $S$ always stays the same, the codimension of 
the critical points of the change-of-variable procedure of \TWO\
increases with the dimension.

\The{\highd}{\Thsty Let $d\ge 3$, and assume \AOne{2,\hxp}, \ATwo{2,\hxp},
\AThr, and, in the asymmetric case, \AFou. Let $G$ be a two--legged
graph contributing to $\Si$, $j <0$, and $\cT$ be a tree, rooted
at a fork $\ph$, and compatible to $G$. Then there is $\de >0$
(independent of $G$) such that 
$$
\sli_{J \in \cJ(\cT,j)} \abs{Val(G^J)}_\two \le
\Const M^{\de j} \la_{n_\ph} (j,\ep) 
\EQN\highhyp $$
}
\Proof The proof is a combination of the methods of \TWO, 
combined with Theorem \twononDOL, Lemma \mskel, and Theorem \twoDOL, 
exactly as was done in the proof of Theorem \zweireg.
We therefore leave it to the reader. \endproof

\appendix{\kern-10pt\hphantom}{: The H\" older argument and the proof of Theorem \Salloreg}
\def\eqhead{A.}

\noindent
In this appendix, we prove Theorem \Salloreg. The essence of
the proof of the H\" older statements of Theorem \Kalloreg\
is the same as in the proof in this appendix:
the extra scale decay leaves the freedom to extract H\" older continuity.
We first illustrate the technique using the example of the second order graph.
All other proofs are easy generalizations of that proof.

\Lem{\Asec}{\Lesty Let $0<\ga <1$ and assume \AOne{1,\ga}, 
\ATwo{2,0}, \AThr, and \AFou. Then $\Si_\two \in C^{1,\ga}(\R\times
\cB,\C)$, i.e.,
$$
\abs{\Si_\two}_{1,\ga} \le K(\ga)
\eqn $$
where $K(\ga)$ diverges in the limit $\ga \to 1$.
}

\Proof
As in \TWO, we only have to deal with the sunset graph because
all other contributions are $C^{1,\ga}$ by the properties of
the scale zero effective action (no derivative or difference
affects the propagators on internal lines of these graphs).

We do the scale decomposition as in Section \TWO.3.1. We then
need to bound $\abs{\cdot}_{1,\ga}$ of 
$$
Y_{\ulj}^\pi (p) =  \int \db p_\one \db p_\two\;
C_{j_\one}(z_\one, e(\p_\one))\;
C_{j_\two}(z_\two, e(\p_\two))\;
C_{j_\thr}(\ze_{\pi(3)},e_{\pi(3)})\;
P_{\pi} (p_\one,p_\two,p)
\EQN\IIIYuljpi $$
where $j_\one<j_\two<j_\thr$,
$$
\ze_a=L_a(z_\one,z_\two,z), \qquad
e_a=e(L_a(\p_\one,\p_\two,\p))
,\EQN\IIIeazea $$
$L_a$ is given by 
$$
L_a (p_\one,p_\two,p) = \cases{
 p+p_\one-p_\two & if $a=1$\cr
 p-p_\one+p_\two & if $a=2$\cr
-p+p_\one+p_\two & if $a=3$\cr}
,\EQN\IIILadef $$
$\pi$ is a permutation of $\{1,2,3\}$, and
$P_\pi$ is given by a $C^{1,\ga}$ function of all momenta with
bounds uniform in $j_\one,j_\two,j_\thr$. 
It  suffices to prove that the scale 
sum over $\{ (j_\one,j_\two,j_\thr): 0>j_\thr > j_\two > j_\one \ge I\}$
of $\abs{Y_{\ulj}^\pi}_{1,\ga}$ converges uniformly in $I$ 
to prove the Lemma (see Section \TWO.3.2).
The convergence of the scale sum for $\abs{Y_{\ulj}^\pi}_{1,0}$ 
was shown in Section \TWO.3.2.

The derivative with respect to $p$ can act only on the product
$C_{j_\thr} P_\pi$, so it gives two terms
$$
\frac{\del}{\del p} Y_{\ulj}^\pi (p) = U_{\ulj}^\pi (p) +
V_{\ulj}^\pi (p) 
\eqn $$
with 
$$\eqalign{
U_{\ulj}^\pi (p) & = \int \db p_\one \db p_\two\;
C_{j_\one}(z_\one, e(\p_\one))\;
C_{j_\two}(z_\two, e(\p_\two))\;
C_{j_\thr}(\ze_{\pi(3)},e_{\pi(3)})\;
\frac{\del P_{\pi}}{\del p} (p_\one,p_\two,p)
\cr
V_{\ulj}^\pi (p) & = \int \db p_\one \db p_\two\;
C_{j_\one}(z_\one, e(\p_\one))\;
C_{j_\two}(z_\two, e(\p_\two))\;
\frac{\del C_{j_\thr}}{\del p}(\ze_{\pi(3)},e_{\pi(3)})\;
P_{\pi} (p_\one,p_\two,p)
\cr}\eqn $$
Both $U_{\ulj}^\pi $ and $V_{\ulj}^\pi$ are $C^{0,\ga}$ functions
of $p$. Since $U_{\ulj}^\pi$ has the same scaling behaviour as
the undifferentiated function, the bound for $\abs{U_{\ulj}^\pi}_{0,\ga}$
is a power of $M^{j_\one}$ better than that for $\abs{V_{\ulj}^\pi}_{0,\ga} $
so we do only the latter bound in detail. 

By \queq{\TWO.3.31}, with $s=1$, 
$$
\abs{V_{\ulj}^\pi}_\zer \le \Const M^{j_\one+j_\two-j_\thr} \abs{j_\thr}
.\eqn $$
Summing over $j_\thr > j_\two$ and $j_\two > j_\one$, we
get
$$
\sli_{j_\two,j_\thr: j_\thr > j_\two > j_\one}
\abs{V_{\ulj}^\pi}_\zer \le \Const \abs{j_\one}^2\;M^{j_\one}
\eqn $$
which is summable over $j_\one < 0$. The difference 
$$
\De_{\ulj}^\pi (p,p') = V_{\ulj}^\pi (p) - V_{\ulj}^\pi (p')
\eqn $$
can be bounded trivially by 
$$
\abs{\De_{\ulj}^\pi (p,p')} \le 2\abs{ V_{\ulj}^\pi}_\zer
\eqn $$
so 
$$
\sli_{j_\two,j_\thr: j_\thr > j_\two > j_\one}
\abs{V_{\ulj}^\pi (p) -V_{\ulj}^\pi (p')} \le
\Const \abs{j_\one}^2 M^{j_\one}
\EQN\onestar $$
We now give a bound for $\abs{\De_{\ulj}^\pi (p,p')}$ that depends
nontrivially on $p$ and $p'$. The difference in $\De_{\ulj}^\pi (p,p')$
acts on a product, so we use the discrete product rule \queq{\diffi}
to convert it into a sum of terms in which the difference acts
on only one factor of the product, and bound all these
terms separately. In other words, we write
$$\eqalign{
\frac{\del C_{j_\thr}}{\del p}(p)\; P_\pi(p_\one,p_\two,p) -
\frac{\del C_{j_\thr}}{\del p}(p')\;P_\pi(p_\one,p_\two,p')
&= \left(\frac{\del C_{j_\thr}}{\del p}(p)-\frac{\del C_{j_\thr}}{\del p}(p')
\right) P_\pi(p_\one,p_\two,p)  
\cr &+\frac{\del C_{j_\thr}}{\del p}(p')
\left(P_\pi(p_\one,p_\two,p) - P_\pi(p_\one,p_\two,p')\right)
\cr}\eqn $$
Consequently, 
$\De_{\ulj}^\pi (p,p')=\De_{\ulj,1}^\pi (p,p')+\De_{\ulj,2}^\pi (p,p')$.
By our hypotheses, $P_\pi$ is $C^{1,\ga}$ with bounds uniform
in the scales, so 
$$
\abs{\De_{\ulj,2}^\pi (p,p')} \le \Const \abs{p-p'} M^{j_\one+j_\two-j_\thr}
\abs{j_\thr}
\eqn $$
which gives a convergent scale sum, as above. 
To bound $\De_{\ulj,1}^\pi (p,p')$, we use 
$$
\frac{\del}{\del p_\al} C_{j_\thr} (\ze_{\pi(3)},e_{\pi(3)})
=  \pm \cases{ (\del_\one C_{j_\thr})(\ze_{\pi(3)},e_{\pi(3)})
& if $\al=0$ \cr
(\del_\two C_{j_\thr}) (\ze_{\pi(3)},e_{\pi(3)}) \;
\frac{\del e}{\del p_\al} (L_{\pi(3)} (p_\one,p_\two,p))
& if $\al \in \nat{d}$,}
\eqn $$
where the sign $\pm$ depends on the sign factor in front of $p$
in $L_{\pi(3)}$. 

Taking the difference of $C_{j_\thr}$ at $p$ and $p'$, we get
again two terms if $\al \in \nat{d}$. The one involving 
$\sfrac{\del e}{\del p_\al} (L_{\pi(3)} (p_\one,p_\two,p)) -
\sfrac{\del e}{\del p_\al} (L_{\pi(3)} (p_\one,p_\two,p'))$
is similar to $\De_{\ulj,2}^\pi (p,p')$ and therefore leads to
a convergent scale sum. 
In the remaining term, we Taylor expand, for $i\in \{1,2\}$, 
the difference (we abbreviate $\ze_{\pi(3)} (p_\one,p_\two,p)=
\ze_{\pi(3)}(p)$ and $e_{\pi(3)} (p_\one,p_\two,p)=e_{\pi(3)}(p)$)
$$\eqalign{
(\del_i C_{j_\thr})&(\ze_{\pi(3)}(p),e_{\pi(3)}(p)) - (\del_i C_{j_\thr})(\ze_{\pi(3)}(p'),e_{\pi(3)}(p'))
\cr &=
\pm \sli_{k=1}^2 \ili_0^1 dt\; 
(\del_k\del_i C_{j_\thr})\left(\ze_{\pi(3)}(p(t)),e_{\pi(3)}(p(t))\right)
\;
\left( (p_\zer-p_\zer')\de_{k1} + (\p-\p') \cdot \nabla e(\p(t)) \de_{k2}\right)
\cr}\eqn $$
with $p(t) = (1-t)p'+tp$. This gives 
$$\eqalign{
\big\vert(\del_i C_{j_\thr})&(\ze_{\pi(3)}(p),e_{\pi(3)}(p)) - 
(\del_i C_{j_\thr})(\ze_{\pi(3)}(p'),e_{\pi(3)}(p'))\big\vert
\cr &\le
(1+\abs{e}_\one) \; \abs{p-p'}\;\ili_0^1 dt\; \max\limits_{i,k\in \{1,2\}}
\abs{\del_k\del_i C_{j_\thr}\left(\ze_{\pi(3)}(p(t)),e_{\pi(3)}(p(t))\right)}
\cr
& \le W_\two (1+\abs{e}_\one) \; \abs{p-p'}\;\ili_0^1 dt\;
M^{-3j_\thr} \True{\abs{e(L_{\pi(3)}(p_\one,p_\two,p(t))} \le
M^{j_\thr}}
\cr}\EQN\Asxtn $$
Inserting this, we get 
$$\eqalign{
\abs{\De_{\ulj,1}^\pi (p,p')} &\le \abs{P_\pi}_\zer (1+\abs{e}_\one)
W_\zer^2 W_\two\;
\abs{p-p'} M^{j_\one+j_\two-3j_\thr} \cr
& \ili_0^1 dt \int d\th_\one d\th_\two
\True{\abs{e(L_{\pi(3)}((0,\p(0,\th_\one)),(0,\p(0,\th_\two)),p(t)))} \le \left(1+2\sfrac{|e|_\one}{u_\zer}\right)M^{j_\thr}}
.\cr}\eqn $$
Since $p_\one$ and $p_\two$ are independent of $t$, we may use
Theorem \bestvol\ to bound the $\th$--integral by 
$\cW((1+2\sfrac{|e|_\one}{u_\zer}))M^{j_\thr}|j_\thr|$, 
which is independent
of $t$. Thus
$$
\abs{\De_{\ulj,1}^\pi (p,p')} \le \abs{P_\pi}_\zer \;
\K(Q_V,\abs{e}_\two,u_\zer)\;
\abs{p-p'}\; \abs{j_\one}\; M^{j_\one+j_\two-2j_\thr} 
\eqn $$
Summing over $j_\thr$ and $j_\two$, we get 
$$
\sli_{j_\two,j_\thr: j_\thr > j_\two > j_\one} 
\abs{\De_{\ulj,1}^\pi (p,p')} \le \Const \abs{p-p'}\; \abs{j_\one}
.\EQN\notco $$
This bound is linear in $p-p'$, but its sum over $j_\one < 0$
diverges. However, taking the weighted geometric mean with
\queq{\onestar}, and adding the terms $\De_{\ulj,2}^\pi (p,p')$
which we estimated before (and which have a convergent scale
sum), we get for any $\ga \in (0,1)$
$$
\sli_{j_\two,j_\thr: j_\thr > j_\two > j_\one} 
\abs{V_{\ulj}^\pi (p) - V_{\ulj}^\pi (p')} \le 
\Const \abs{p-p'}^\ga \; M^{j_\one (1-\ga)} \abs{j_\one}^{2-\ga}
.\eqn $$
The sum over $j_\one$ now converges, which proves the statement
of the Lemma. \endproof

To show the same for $\Si_r$, for all $r \ge 3$, it suffices
to bound the value of any order $r$ two-legged 1PI graph. 
Thus Theorem \Salloreg\ follows from the next Lemma.

\Lem{\Sallolem}{\Lesty Let $G$ be a 1PI two--legged graph and
$J$ a labelling of $G$. Let $0<\ga<1$. Assume \AOne{1,\ga},
\ATwo{2,0},\AThr, and \AFou. Let $\cT$ be a tree consistent with
$G$. Then for all $j<0$, and all labellings $J \in \cJ(\cT,j)$
of $G$ ($\cJ$ is defined in \queq{(\ONE.2.73)}), $Val(G^J)$ is
$C^{1,\ga}$, and there is $\al >0$ and a constant $K$ such that 
for all $\si \in \natz{d}$
$$
\sli_{J \in \cJ(\cT,j)} \abs{\frac{\del}{\del p_\si}Val(G^J) (p)
- \frac{\del}{\del p_\si}Val(G^J) (p')} \le 
K \abs{j}^\al M^{j(1-\ga)}\abs{p-p'}^\ga
.\EQN\Sallox $$
$K$ and $\al $ depend on $G$. For any two--legged 1PI graph
$G$, the sum 
$$
\sli_{j<0} \sli_{\cT\sim G}\sli_{J \in \cJ(\cT,j)} \abs{Val(G^J)}_{1,\ga} 
\EQN\Salloxx $$
converges.
}

\Proof 
It suffices to prove \queq{\Sallox} because \queq{\Salloxx} 
then follows simply by summation over $\cT$ and $j$.
By \AOne{1,\ga} and Lemma \TWO.2.3, the vertex functions
associated to the scale zero effective vertices are $C^{1,\ga}$,
with bounds uniform in the scales. The bounds depend only on
$\abs{\hat v}_{1,\ga}$. We do an induction in the depth of $(\cT,G)$,
defined as 
$$
P_\two (G,\cT) = \max\{ k: \exists f_\one > f_\two > \ldots > f_k
>\ph: E(G_{f_l})=2 \hbox{ for all } l \in \nat{k}\}
\eqn $$
with \queq{\Sallox} as the inductive hypothesis.
If $P_\two=0$, $G$ is a skeleton graph (or $G$ has at most one
vertex, in which case the statement is obvious by the properties
of the vertex functions). 

Let $\tG(\ph)$ be the root scale quotient of $G$ (see Definition
\ONE.2.27). Let $P_\two(G,\cT)=0$.
\ssni
{\it Case 1:} $\tG(\ph)$ is \OL. We take the difference 
$$
\De_J(p,p')=\left(\frac{\del}{\del p_\si}Val(G^J)\right) (p) -
\left(\frac{\del}{\del p_\si}Val(G^J)\right) (p')
.\eqn $$
As in second order, the derivative can act on vertex functions
or propagators. Again, we rewrite the result by the discrete
product rule \queq{\diffi}. This gives a sum of terms of the
following types (denoted by $T^{(i)}_J$)

\leftit{(1)}  Both the derivative and the difference operator
act on a vertex function.
This term has the same scale behaviour as the undifferentiated
graph because the vertex function has uniform bounds. Since 
$\tG(\ph)$ is \OL, Theorem \ONE.2.46 $(i)$ implies that 
$$
\sli_{J \in \cJ(\cT,j)} \abs{T^{(1)}_J(p,p')} \le 
\Const \abs{j}^{\al_\one} M^{2j} \abs{p-p'}
.\EQN\IIIo $$

\leftit{(2)} A vertex function gets differentiated, but 
the difference operator acts on a propagator. 
Rewriting the difference of the propagator
at $p$ and $p'$ by Taylor expansion, we see that this term behaves
like the first derivative of the value of the graph. By Theorem
\ONE.2.46 $(i)$, 
$$
\sli_{J \in \cJ(\cT,j)} \abs{T^{(2)}_J(p,p')} \le 
\Const \abs{j}^{\al_\two} M^{j} \abs{p-p'}
\EQN\IIIx $$

\leftit{(3)} The derivative acts on the propagator on line $\ell_\one$
and the difference on line $\ell_\two$ ($\ell_\one=
\ell_\two$ is possible). With the same Taylor expansion as above,
the scale behaviour deteriorates by a factor $\Const M^{-j_{\ell_\one}
-j_{\ell_\two}} \le M^{-2j}$. Thus this term is bounded by 
$$
\sli_{J \in \cJ(\cT,j)} \abs{T^{(2)}_J(p,p')} \le 
\Const \abs{j}^{\al_\thr} \abs{p-p'}
.\EQN\IIIxx $$
\ssni
Summing up all these terms, we obtain
$$
\sli_{J \in \cJ(\cT,j)} \abs{\De_J(p,p')} \le
\Const \abs{j}^\al \abs{p-p'}
.\EQN\IIIxxx $$
On the other hand, we also have the trivial estimate
$$
\sli_{J \in \cJ(\cT,j)} \abs{\De_J(p,p')} \le
2 \sli_{J \in \cJ(\cT,j)} \abs{\frac{\del}{\del p_\si}Val(G^J)}_\zer
\le \Const M^{j} \abs{j}^\al 
\EQN\IIIxxxx $$
The geometric mean of \queq{\IIIxxx} and \queq{\IIIxxxx} gives
$$
\sli_{J \in \cJ(\cT,j)} \abs{\De_J(p,p')} \le
\Const \abs{p-p'}^\ga \; M^{j(1-\ga)} \abs{j}^\al
.\EQN\IIIyeah $$

\ssni
{\it Case 2:} $\tG(\ph)$ is \NOL. Since $P_\two(G,\cT)=0$, this
means that $\tG(\ph)$ is an ST diagram (see Definition \ONE.2.21).
Thus all lines on scale $j$ are self--contractions, and therefore
the propagators associated to them do not depend on the external
momentum $p$. Thus no derivative or difference operation can
act on these propagators. Let $\ta_\ph$ be the maximal
subtree of $t$ rooted at $\ph$ such that $\tG(\ta_\ph)$ is \NOL.
By Lemma \ONE.2.31 $(ii)$, $\ta_\ph$ exists and is unique. 
Let $\tG=\tG(\ta_\ph)$. By definition of $\ta_\ph$,  
the lowest scale $j^*$ on which
the derivative or difference can have an effect is also the scale
where a volume gain from overlapping loops occurs (if $j^*=0$,
there is no volume gain, but then the derivative acts only on
a scale zero vertex function, and thus does not affect the scale
behaviour at all). All considerations of the previous case apply,
only with $j$ replaced by $j^*$. Thus the derivatives and differences
produce at worst a factor $M^{-2j^*}$ and volume improvement
produces a factor $M^{j^*}$. Summing the scales, 
we get again \queq{\IIIxxx} and \queq{\IIIxxxx}. 

Let $P_\two (G,\cT) >0$. Now $G$ can have two--legged subgraphs,
corresponding to $r$-- and $c$--forks of $t$. They have smaller
depth $P_\two$. Therefore the inductive hypothesis applies to
them. We construct the skeleton graph $G'$ associated to $G$
by replacing the strings of two--legged subdiagrams by new propagators.
We now verify that these strings have the same 
behaviour as ordinary propagators when they are differentiated
and when differences are taken. 

The value of such a string is given by \queq{(\ONE.2.93)}.
When the derivative and the difference operator act on one 
of the propagators the scaling behaviour changes in the same
way as for a single propagator. If they act on an $r$--fork,
the renormalization cancellation is lost, but the inductive hypothesis
implies that no other factor $M^{-\be j}$, $\be >0$, occurs.
If the derivative acts on a $c$--fork, by \queq{\IIIx}, the net
effect is the same as if the derivative had acted on a line of
scale $j_\ph$ rather than internal to the $c$--fork. 
If the derivative and the difference operator
act on the $c$--fork, the inductive hypothesis
\queq{\IIIyeah} implies that the $c$--fork
scale sum is still convergent. If derivatives and differences
act on different factors in the value of the string, 
Taylor expansion implies that the same bounds hold.
Thus strings behave like single propagators, which
puts us back to the case $P_\two=0$, which we have already done.
\endproof

Finally, we state the Lemma that implies that the strings of
two--legged subdiagrams $S_{l,j_l}$ 
have the properties required of the general
propagators appearing in Lemma \TWO.3.8.

\Lem{\yuck}{\Lesty Assume the hypotheses of Lemma $\stringprop$,
assume \ATwo{2,\hxp}, and
and assume that there are $\ta_{k,\hxp}>0$ and $n_k \in \N$ such that 
$$
\abs{D^\al T_k^{(j)} (p) - D^\al T_k^{(j)}(p')} \le 
\ta_{k,\hxp} \abs{p-p'}^\hxp \abs{j}^2 M^{-\hxp j} \la_{n_k} (j,\veps)
\EQN\IIIyum $$
for all multiindices $\al$ with $|\al|=2$, for some $\veps >0$.
Then 
$$
\abs{D^\al S_{l,j_l}^{(\nu_l)} (p) - D^\al S_{l,j_l}^{(\nu_l)} (p')}
\le H_S \abs{p-p'}^\be \; 
M^{-j_l (\nu_l+k-1+\be)} \; 
\True{\abs{ip_\zer - e(\p)} \le G_\zer M^{j_l}}
\eqno{(\TWO.3.126)} $$
and 
$$\eqalign{
& \abs{\left(\sfrac{\del}{\del \th}\right)^{k-1} S_{l,j_l}^{(\nu_l)} (p_\zer,
\p(\rh,\th)) - \left(\sfrac{\del}{\del \th}\right)^{k-1} S_{l,j_l}^{(\nu_l)} (p_\zer,
\p(\rh',\th'))} \cr
& \qquad \le H_S \abs{\p(\rh,\th)-\p(\rh',\th')}^\be \; 
M^{-j_l (\nu_l + \be)} \; 
\True{\abs{ip_\zer - \rh} \le 4 M^{j_l}}
\cr}\eqno{(\TWO.3.127)} $$
hold, with $\nu_l=1$, $\be=\hxp$ and 
$H_s = \Const \abs{j}^2 \prod_k \la_{n_k}(j,\veps)$
where the constant depends only on $\ta_{1,\hxp}, \ldots, \ta_{n,\hxp}$
and $\abs{e}_{2,\hxp}$ and $\veps$. 
}

\Proof If $\abs{p-p'} \ge M^j$, then $\abs{p-p'}^{-\be} \le M^{-j\be}$
so the statement follows from $|S_j(p)-S_j(p')| \le |S_j(p)|+|S_j(p')|$,
and Lemma \stringprop. So let $|p-p'| < M^j$. 
Using the discrete product rule in the usual way, we get 
contributions involving the differences 
$D^{\al'} T_k^{(j)}(p) - D^{\al'} T_k^{(j)}(p') $ and
$D^{\al'} C_j(p_\zer,e(\p)) - D^{\al'} C_j(p_\zer',e(\p'))$
with $\al' \le 2$. If $|\al'| <2$, Taylor expansion and the inequality
$$
\abs{p-p'} \le \abs{p-p'}^\hxp M^{j(1-\hxp)}
\EQN\jojo $$
do the job. 
In particular, we have for propagators an obvious analogue of
\queq{\Asxtn} with, possibly, a more complicated linear combination
of other momenta, and \queq{\jojo} implies a bound involving
$M^{-j(1+\hxp+\abs{\al'})}$.
Let $\al'=2$. If the difference acts on $D^{\al'} T_k^{(j)}$, 
the result follows from \queq{\IIIyum}. 
$D^{\al'} C_j$ contains the summand $(\del_\two C_j) D^{\al'}e$,
which is the only summand containing the second derivative of
$e$. In the difference, we use again the discrete product rule. 
The term involving the difference of $e $ is bounded as
$$
\abs{D^{\al'}e(p) - D^{\al'}e(p')} \le \abs{e}_{2,\hxp} \; \abs{p-p'}^\hxp
.\eqn $$
In the other terms, we are back to the case (with $\al' \le 1$).
\endproof

\Rem{\machtnix} The polynomial dependence of $H_S$ on $j$ does
not change the conclusion of Theorem \TWO.3.8 because the convergence
factor in the scale sum is $M^{\de j}$ with $\de >0$. 

\vfill\eject
\centerline{\bfe References}
\sni
{\parskip=0.5cm
\Ref{F}{S.\ Fujimoto, {\sl Anomalous Damping of Quasiparticles
in Two--Dimensional Fermi Systems, Journal of the Physical Society
of Japan \bf 60}(1991) 2013}
\Ref{FST1}{J.\ Feldman, M.\ Salmhofer and E.\ Trubowitz, 
{\sl Perturbation Theory around Non-Nested Fermi Surfaces I.
Keeping the Fermi Surface Fixed}, 
{\sl J.\ Stat.\ Phys.\bf 84} (1996) 1209}
\Ref{FST2}{J.\ Feldman, M.\ Salmhofer and E.\ Trubowitz, 
{\sl Perturbation Theory around Non-Nested Fermi Surfaces II.
Regularity of the Moving Fermi Surface: RPA graphs}, mp-arc 96-684,
submitted to {\sl Communications in Pure and Applied Mathematics}}
\Ref{FT1}{J.Feldman and E.Trubowitz, {\sl Perturbation Theory for
Many--Fermion Systems, Helvetica Physica Acta \bf 63} (1990) 156}
\Ref{FT2}{J.Feldman and E.Trubowitz, {\sl The Flow of an Electron--Phonon
System to the Superconducting State, Helvetica Physica Acta \bf 64}
(1991) 213}
\Ref{GN}{G.\ Gallavotti and F.\ Nicol\` o, {\sl Renormalization Theory
in Four Dimensional Scalar Fields}, {\it Comm.\ Math.\ Phys.\ \bf 100} (1985)
545 and {\bf 101} (1985) 247}
\Ref{S}{M.\ Salmhofer, {\sl Improved Power Counting and Fermi Surface 
Renormalization}, cond-mat/9607022}
\par}

\end

%% file: mnsmac.tex
%
%
%
%
\immediate\openout8=\jobname.txs
\newcount\sectno
\newcount\chapno
\newcount\equano
\newcount\theono
\newcount\refno
\sectno=0
\chapno=0
\equano=0
\theono=0
\refno=0
\def\eqhead{ }
\def\\{\backslash}
\newif\ifintrmk
\def\irmkanf{\vglue 8pt\noindent\hbox to\hsize{\hrulefill}\pni}
\def\irmkend{\pni\hbox to\hsize{\hrulefill}\vglue 8pt\noindent}

\font\bfe=cmbx12
\def\chapskip{\removelastskip\par\vglue 40pt}
\def\sectskip{\removelastskip\par\vskip 40pt}
\def\blwskip{\removelastskip\par\vskip 20pt}
\def\chap#1{\equano=0 \sectno=0 \theono=0 \global\advance\chapno by 1%
\def\eqhead{\number\chapno .}%
\chapskip\goodbreak\centerline{\bfe \eqhead \hglue 5pt #1}
\blwskip}%
\def\sect#1{\global\advance\sectno by 1
\sectskip\goodbreak%
\noindent{\bf \eqhead\number\sectno  \hglue 5pt #1}
\blwskip}%
\def\appendix#1#2{\par\chapskip\goodbreak\centerline{\bfe Appendix #1. #2}%
\blwskip%
\equano=0\sectno=0\theono=0\def\eqhead{#1.}}
%
%
%
%
%
%
%
\edef\Raum{ }
\def\eqn{{\hbox{\global\advance\equano by 1}}%
\eqno (\eqhead\number\equano )}%
\def\EQN#1{\eqn\edef\Zwi{(\eqhead\number\equano )}%
\immediate\write8{EQN  \Zwi\Raum = \noexpand#1}
\global\let #1=\Zwi
}
\def\queq#1{$#1$}
\def\The#1{{\global\advance \theono by 1 \someroom \noindent%
{\bf Theorem
\eqhead\number\theono }\hroom
\edef\Zwi{\eqhead\number\theono}
\immediate\write8{STM  \Zwi\Raum = \noexpand#1    Theorem }
\global\let#1=\Zwi
}}
\def\Pro#1{{\global\advance \theono by 1 \someroom \noindent%
{\bf Proposition
\eqhead\number\theono }\hroom
\edef\Zwi{\eqhead\number\theono}
\immediate\write8{STM  \Zwi\Raum = \noexpand#1    Proposition }
\global\let#1=\Zwi}}
\def\Rem#1{{\global\advance \theono by 1 \someroom \noindent%
{\bf Remark
\eqhead\number\theono }\hroom
\edef\Zwi{\eqhead\number\theono}
\immediate\write8{STM  \Zwi\Raum = \noexpand#1    Remark }
\global\let#1=\Zwi}}
\def\Cor#1{{\global\advance \theono by 1 \someroom \noindent%
{\bf Corollary
\eqhead\number\theono }\hroom
\edef\Zwi{\eqhead\number\theono}
\immediate\write8{STM  \Zwi\Raum = \noexpand#1    Corollary }
\global\let#1=\Zwi}}
\def\Def#1{{\global\advance \theono by 1 \someroom \noindent%
{\bf Definition
\eqhead\number\theono }\hroom
\edef\Zwi{\eqhead\number\theono}
\immediate\write8{STM  \Zwi\Raum = \noexpand#1    Definition }
\global\let#1=\Zwi}}
\def\Lem#1{{\global\advance \theono by 1 \someroom \noindent%
{\bf Lemma
\eqhead\number\theono }\hroom
\edef\Zwi{\eqhead\number\theono}
\immediate\write8{STM  \Zwi \Raum = \noexpand#1    Lemma }
\global\let#1=\Zwi
}}
\def\Proof{\someroom \noindent{\it Proof:}\hroom}
\def\endproof{\hfill
\hbox{\vrule width 7pt depth 0pt height 7pt} \vskip 25pt plus
5pt minus 10pt}

\def\refit#1{\par\noindent\hangindent=1.5cm\hangafter=1
           \hbox to 1cm{#1\hss}\hglue 0.5cm\ignorespaces}
\def\Ref#1#2{\refit{[#1]} #2}
\def\quref#1{[#1]}
\def\today{\ifcase\month\or
 January\or February\or March\or April\or May\or June\or
 July\or August\or September\or October\or November\or December\fi
 \space\number\day, \number\year}
\def\getArchdate{\openin7=\jobname.arx
                 \ifeof7 \def\Archdate{\today}
                 \else \read7 to \Archdate\fi
                 \closein7}
%
%
\newcount\figno
\figno=0
\def\nextfig#1{\global\advance\figno by 1%
\edef\Zwi{\number\figno}%
\immediate\write8{FIG  \Zwi\Raum = \noexpand#1   }%
\global\let#1=\Zwi $\Zwi$}
%
%
\def\senzafig#1{\global\advance\figno by 1%
\edef\Zwi{\number\figno}%
\immediate\write8{FIG  \Zwi\Raum = \noexpand#1   }%
\global\let#1=\Zwi}

%
%
%
%
\def\N{{\rm I\kern-.16em N}}
\font\twblk=cmss10
\font\tenblk=cmss8
\font\eiblk=cmss8
\def\Z{\mathchoice{{\hbox{\twblk Z\kern-.35emZ}}}
{{\hbox{\twblk Z\kern-.35emZ}}}
{{\hbox{\tenblk Z\kern-.30emZ}}}
{{\hbox{\eiblk Z\kern-.24emZ}}}}
\def\Q{{\rm \kern.25em\vrule height1.4ex
depth-.12ex width.06em\kern-.31em Q}}
\def\R{{\rm I\kern-.2emR}}
\def\C{{\rm \kern.25em\vrule height1.4ex
depth-.12ex width.06em\kern-.31em C}}
\def\K{{\rm I\kern-.2emK}}
\def\L{{\rm I\kern-.2emL}}
\def\bbbone{{\mathchoice {\rm 1\mskip-4mu l} {\rm 1\mskip-4mu l}    
{\rm 1\mskip-4.5mu l} {\rm 1\mskip-5mu l}}}
%
%
\def\al{\alpha}
\def\be{\beta}
\def\ga{\gamma}
\def\de{\delta}
\def\ep{\epsilon}
\def\veps{\varepsilon}

\def\ze{\zeta}
\def\th{\theta}
\def\vth{\vartheta}

\def\ka{\kappa}
\def\la{\lambda}
\def\rh{\rho}
\def\si{\sigma}
\def\ta{\tau}

\def\ph{\phi}

\def\ch{\chi}
\def\ps{\psi}

\def\veps{\varepsilon}
%
%
\def\Ga{\Gamma}
\def\De{\Delta}

\def\La{\Lambda}
\def\Si{\Sigma}

\def\Om{\Omega}
%
%

\def\chq{{\bar\chi}}
\def\psq{{\bar\psi}}

%

%
%

%
%

\def\cB{{\cal B}}
\def\cC{{\cal C}}

\def\cE{{\cal E}}
\def\cF{{\cal F}}
\def\cG{{\cal G}}

\def\cJ{{\cal J}}

\def\cM{{\cal M}}

\def\cS{{\cal S}}
\def\cT{{\cal T}}
\def\cU{{\cal U}}
\def\cV{{\cal V}}
\def\cW{{\cal W}}

%
%
\def\zer{{\oldstyle 0}}
\def\one{{\oldstyle 1}}
\def\two{{\oldstyle 2}}
\def\thr{{\oldstyle 3}}
\def\tG{{\tilde G}}

\def\db{{\mkern2mu\mathchar'26\mkern-2mu\mkern-9mud}}

\def\I2{{$I_2$}}
\def\del{\partial}

\def\ve#1{{\bf #1}}

\def\tst#1{{{\theta_{#1}}^*}}

\def\nat#1{\{ 1,\ldots,#1 \} }
\def\natz#1{\{ 0,\ldots,#1 \} }
\def\abs#1{{\left\vert #1 \right\vert}}

\def\openkrnl#1{\mathop{#1}\limits^{\;_\zer}}
%
%
\def\True#1{\; \bbbone\left( #1 \right) \;}
%
%
         
%
%
         
%
\def\dst{\displaystyle}

\def\tst{\textstyle}
\def\frac#1#2{\dst {#1\over#2}}     
\def\sfrac#1#2{{\tst{#1\over#2}}}   

\def\dotcup{\mathop{{\mathop{\cup}\limits^\cdot}}}
\def\Const{\hbox{ \rm const }}
\def\supp{\hbox{ \rm supp }}
\def\NOL{non--\-over\-lap\-ping}
\def\OL{{o\-ver\-lap\-ping}}

\def\ili{\int\limits}
\def\sli{\sum\limits}
\def\pli{\prod\limits}

\def\suffix{ps}
\newcount\system
\global\system=3   

\def\ifundefined#1{\expandafter\ifx\csname#1\endcsname\relax}
\ifundefined{figdir}\def\figdir{}\fi
%
%
\newcount\firstline
\newdimen\pswidth  \newdimen\xleft
\newdimen\psheight \newdimen\ytop \newdimen\ybot
\newcount\justx \newcount\justy
\global\justx=0 \global\justy=0
\newdimen\vpos \newtoks\label 
\newread\labelfile \newdimen\xcoord \newdimen\ycoord
\newif\ifdoit 
\newbox\labox
\newdimen\xdvikwid 
\newdimen\xdvikht
\newdimen\pspoints
\newdimen\rwi
\pspoints=1bp
\newcount\temp
\def\readdim#1{\global\read\labelfile to \temp
\global #1=\temp pt}
%
%
%
%
\def\figcrop#1{\par
\openin\labelfile=\figdir#1.lbl                                              
\global\read\labelfile to\firstline\message{#1}               
\global\read\labelfile to\temp
\readdim{\ybot}
\readdim{\xleft}
\readdim{\ytop}
\global\read\labelfile to\justx
\global\read\labelfile to\justy
\global\read\labelfile to\label
\readdim{\pswidth}
\global\advance\pswidth by -\xleft
\readdim{\psheight}
\global\advance\ybot by -\psheight
\global\advance\psheight by -\ytop
\global\read\labelfile to\justx
\global\read\labelfile to\justy
\global\read\labelfile to\label
\vbox to\psheight{\vfill
\ifnum\system=1
\fi 
\ifnum\system=2
\fi 
\ifnum\system=3
\fi         
\ifnum\system=4
\fi         
\ifnum\system=1
\hbox to \pswidth{\kern-\xleft\special{postscriptfile \figdir#1.\suffix }\hfil}\fi
\ifnum\system=2
\hbox to \pswidth{\kern-\xleft\special{ps: plotfile \figdir#1.\suffix }\hfil}\fi
\ifnum\system=3
\hbox to \pswidth{\kern-\xleft\includegraphics{\figdir#1.\suffix}\hfil}\fi
\ifnum\system=4
\hbox to \pswidth{\kern-\xleft\includegraphics{\figdir#1.\suffix}\hfil}\fi
\ifnum\system=5
\hbox to \pswidth{\kern-\xleft\includegraphics{\figdir#1.\suffix}\hfil}\fi 
\ifnum\system=6
   \xdvikwid=\pswidth
   \xdvikht=\psheight
   {\global\divide\xdvikwid by \pspoints}
   {\global\divide\xdvikht by \pspoints}
   \rwi=\xdvikwid
    {\global\multiply\rwi by 10}
\hbox to \pswidth{\kern-\xleft\includegraphics{\figdir#1.\suffix\space}\hfil}\fi                   
\vskip -\baselineskip
\vskip -\ybot 
\vskip-\psheight %
\hbox to\pswidth  {\hss}%
\parindent=0pt\offinterlineskip                                       
\vpos=0 pt%
\loop\readdim{\xcoord}                                 
\ifdim \xcoord < -999pt \doitfalse\else\doittrue\fi                        
\ifdoit \advance \xcoord by -\xleft
\readdim{\ycoord}
\advance \ycoord by -\ytop                              
\global\read\labelfile to\justx                                       
\global\read\labelfile to\justy                                       
\global\read\labelfile to\label
\global\setbox\labox=\hbox{\label\hskip-0.3em}%
\advance\vpos by-\ycoord                                              
\vskip-\vpos \vpos=\ycoord                                         
\hbox to\pswidth{\hskip\xcoord %
\hbox to 0pt{\ifnum\justx>0\hss\fi%
\vbox to0pt{%
\ifnum\justy<2\vss\fi%
\copy\labox\kern0pt%
\ifnum\justy>0\vss\fi}%
\ifnum\justx<2\hss\fi}%
\hss}%
\repeat%
\advance\vpos by-\psheight%
\vskip-\vpos %
}\closein\labelfile}
%
%
%
\def\figplace#1#2#3{
\openin\labelfile=\figdir#1.lbl
\ifeof\labelfile\immediate\write16{Can't find \figdir#1.lbl; I quit!}\end\fi 
\closein\labelfile
\null\hskip#2\raise #3 \hbox{\figcrop{#1}}
}
%
%
\def\herefig#1{%
\ifshowfigs
\midinsert
\centerline{\figplace{#1}{0in}{0in}}
\endinsert
\fi
}
%
%
\ifintrmk
\message{Internal remarks will be shown}
\else
\message{To see internal remarks, change intrmkfalse to intrmktrue}
\fi
\getArchdate
\message{File last changed on \Archdate}